\documentclass{aa}

\usepackage{graphicx}
\usepackage{txfonts}
\usepackage{bm}
\usepackage{hyperref}
\usepackage{comment}
\begin{document}

   \title{Characterizing Lyman alpha emission from high-redshift galaxies}

   \author{Samuel Gagnon-Hartman\thanks{email:samuel.gagnonhartman@sns.it}\inst{1}
          \and
          Andrei Mesinger\inst{2, 5, 1}
          \and
          Ivan Nikoli\'c\inst{3,4}
          \and
          Eleonora Parlanti\inst{1}
          \and
          Giacomo Venturi\inst{1}
          }

   \institute{Scuola Normale Superiore di Pisa,
              Piazza dei Cavallieri 7, 56126 Pisa, Italy
             \and
             Department of Physics and Astronomy {\it ``Ettore Majorana''}, University of Catania, Via Santa Sofia 64, 95123  Catania, Italy
             \and
             Cosmic Dawn Center (DAWN)
             \and
             Niels Bohr Institute, University of Copenhagen, Jagtvej 128, 2200 Copenhagen N, Denmark
             \and
            Centro Nazionale ``High Performance Computing, Big Data and Quantum Computing'', Via Magnanelli, 40033 Casalecchio di
             Reno, Italy\\
            }

   \date{Received February 6, 2026; accepted February 6, 2026}
 
\abstract{  
The Lyman $\alpha$ (Ly$\alpha$) line from high-redshift galaxies is a powerful probe of the Epoch of Reionization (EoR).  Neutral hydrogen in the intergalactic medium (IGM) can significantly attenuate the emergent Ly$\alpha$ line, even in the damping wing of the cross-section.  However, interpreting this damping wing imprint relies on our prior knowledge of the spectrum that escapes from the galaxy and its environs into the IGM.  This emergent spectrum is highly sensitive to the composition and geometry of the interstellar and circumgalactic media, and so exhibits a large galaxy to galaxy scatter.  Characterizing this scatter is further complicated by non-trivial selection effects introduced by observational surveys.  
Here we build a flexible, empirical model for the emergent Ly$\alpha$ spectra.  Our model characterizes the emergent Ly$\alpha$ luminosity, the velocity offset of the Ly$\alpha$ line with respect to the systemic redshift, and the H$\alpha$ luminosity, with multivariate probability distributions conditioned on the UV magnitude.  
We constrain these distributions using $z\sim5-6$ galaxy observations with VLT MUSE and {\it JWST} NIRCam, forward-modeling observational selection functions together with galaxy parameters.
Our model results in Ly$\alpha$ equivalent width distributions that are a better match to (independent) Subaru observations than previous empirical models.
The extended distributions of Ly$\alpha$ equivalent widths and velocity offsets we obtain could facilitate Ly$\alpha$ transmission during the early stages of the EoR.
We also illustrate how our model can be used to identify GN-z11-like outliers, potentially originating from merging systems.
We publish fitting functions and make our model publicly available.
}


\maketitle


\section{Introduction}

The first stars, black holes and galaxies formed sometime during the first billion years, starting the so-called Cosmic Dawn (CD). The radiation from these galaxies spread out, heating and ionizing the intergalactic medium (IGM), culminating in the last phase change of our Universe: the Epoch of Reionization (EoR). The details of this process, including when reionization happened and which objects drove it, remain uncertain (see reviews by, e.g., \citealt{mesinger16, dayal18}). While a definitive measurement of the timing and morphology of reionization will likely come from the  redshifted 21-cm line of neutral hydrogen, making this observation is very technically challenging and it might be a long time until we have robust data.

Currently one of the most powerful techniques to study the EoR is through the observed Lyman-$\alpha$ (Ly$\alpha$) spectra of galaxies (e.g., \citealt{dijkstra14, hutter17,mason2018,jung2020,umeda25}). During the EoR there is sufficient neutral hydrogen (HI) in the IGM to significantly attenuate Ly$\alpha$ even via the damping wing of the cross section.  The damping wing attenuation can extend redward of the line center, impacting spectral regions that might otherwise be unaffected by resonant Ly$\alpha$ scattering from HI in the vicinity of the galaxy \citep{dijkstra14,mesinger15,gelli25}.  Identifying the IGM damping wing imprint in the observed Ly$\alpha$ spectra of a galaxy would allow us to measure the average hydrogen neutral fraction in the IGM, $\bar{x}_{\rm HI}$, and even the EoR morphology (given a sufficiently large sample of galaxies; e.g. \citealt{furlanetto06, mcquinn07,mesinger08,stark10,sobbachi15,mason2018,jung2020,bolan22,tingyi24,nikolic24,nakane24,umeda25,jones25}).

Unfortunately, measuring the IGM damping wing imprint from the observed spectra requires knowing the Ly$\alpha$ profile that emerges from the galaxy into the IGM.
Ly$\alpha$ in galaxies is primarily sourced by recombinations in the ionized interstellar medium (ISM) surrounding type-O and B stars \citep{hui98,greif09,raiter10,pawlik11}, although collisionally excited HI in the ISM and circumgalactic medium (CGM) can act as a secondary source \citep{haiman2000,fardal01}. To be detected, the Ly$\alpha$ photons must escape from the galaxy in the direction of the observer. HI and dust in the ISM and CGM resonantly scatter and absorb Ly$\alpha$ photons. The geometry of the ISM and CGM varies from galaxy to galaxy, rendering the relationship between intrinsic Ly$\alpha$ photon production and the emergent Ly$\alpha$ spectrum nontrivial.

Various models have been proposed which connect galaxy properties to properties of the emergent Ly$\alpha$ spectrum. These include analytic models, which make simplifying assumptions about galaxy geometry and composition to predict the emergent spectrum, radiative transfer simulations through the ISM and CGM, which aim to capture the full complexity of Ly$\alpha$ propagation in realistic environments, and empirical models, which draw scaling relations between galaxy and Ly$\alpha$ properties from observations.

The fully analytic solution of \cite{neufeld90} provides a handy relationship between the velocity offset of the Ly$\alpha$ line and the HI column density of the system, but this only holds in the case of a static slab geometry with a singly-peaked Ly$\alpha$ source placed at its center. Works employing numerical radiative transfer have since explored other geometries, most famously the expanding spherical shell and its variants \citep{verhamme12,gronke16,nebrin25,smith25}.
Others combine galaxy  simulations with radiative transfer to produce Ly$\alpha$ spectra not reliant on simple geometries \citep{behrens18,behrens19,byrohl20,smith22}. Recent work by \cite{khoraminezhad25} aimed to bridge the gap between analytic and empirical modeling by fitting a series of empirical scaling relations to the input of \texttt{zELDA}, a Ly$\alpha$ spectrum emulator trained on a database of expanding spherical shell model spectra (\texttt{zELDA}: \citealt{gurunglopez25}). They tuned the parameters of their model to accurately reproduce the luminosity function and clustering of Ly$\alpha$-emitting galaxies at $z\sim2-3$. While such simulations provide qualitative insight into emergent spectra, the inherent assumptions about ISM/CGM geometries can result in non-physical best-fit parameters (e.g. \citealt{li22, gronke17,vitte25}).

Due to these challenges, studies of the EoR typically rely on empirical scaling relations that were fitted to observed Ly$\alpha$ spectra unattenuated by the EoR (i.e. at $z\lesssim6$).  
For example, \cite{mason2018} (M18) proposed a simple model which characterizes the probability of Ly$\alpha$ emission, the mean Ly$\alpha$ equivalent width, and the mean velocity offset as functions of the UV continuum magnitude ${\rm M}_{\rm UV}$ (see also \citealt{treu12,tang24}). Recently, \cite{prietolyon25} expanded on such a model using a larger $z\sim5$ dataset, fitting the Ly$\alpha$ equivalent width PDF as a function of UV magnitude and slope (although they find no dependence on the later).
Such empirical relations are proving to be invaluable in interpreting the ever-increasing sample of Lyman alpha galaxy spectra during the EoR (e.g., \citealt{saxena23,tang23,chen24,napolitano24,tang24b,umeda24b,witstok25}).

However, empirical models also have shortcomings. In the absence of guiding physics, the functional forms and dependencies need to be either chosen a priori or motivated by the data (e.g. using Bayesian model selection, see review by \citealt{trotta08}).  
Whatever the choice of functional model, there is no guarantee that it can be generalized to other datasets.  
This is especially true if the data used to fit the model had inhomogeneous or poorly-understood selection effects. Biased observational samples, as well as poorly-chosen functional models, can yield spurious correlations or hide true correlations (e.g., \citealt{dicesare25}).

Here we build a flexible, data-driven empirical model for the emergent (i.e. leaving the ISM/CGM and entering the IGM) Lyman alpha profile.  Our model characterizes the emergent Ly$\alpha$ luminosity, the velocity offset of the Ly$\alpha$ line with respect to the systemic redshift, and the H$\alpha$ luminosity, with probability distributions conditioned on the UV magnitude.  We also forward model selection effects together with the model parameters. We calibrate our model on the recent homogeneous sample from \cite{tang24}, composed of objects detected in Ly$\alpha$ by VLT MUSE and in H$\alpha$ by {\it JWST} NIRCam (MUSE-Wide: \citealt{urrutia19}; MUSE-Deep: \citealt{bacon17,bacon23}; \textit{JWST} FRESCO: \citealt{oesch23}).
We demonstrate that our model can reproduce high-$z$ Ly$\alpha$ statistics, such as the Ly$\alpha$ luminosity function and equivalent width distribution measured by \cite{umeda25}, despite being calibrated on an independent dataset.

This paper is organized as follows. Section \ref{sec:meth} details the impact of selection effects on inferred probability distributions, describes the \cite{tang24} (T24) galaxy sample and its selection criteria, and then goes on to describe our model and its calibration. A summary of our model is included at the end of the section for ease of implementation. Section \ref{sec:res} lists the correlations found by our model, giving analytic scaling relations for the Ly$\alpha$ luminosity, velocity offset, H$\alpha$ luminosity, Ly$\alpha$ escape fraction, and Ly$\alpha$ equivalent width with ${\rm M}_{\rm UV}$. Section \ref{sec:disc} compares these relations to those found by previous work, and explores some consequences of our model. Section \ref{sec:concl} enumerates our main results. Throughout this work we adopt the cosmology of \cite{planck18}, where $H_0=67.66$ km s$^{-1}$, $\Omega_m=0.30966$, and $\Omega_b=0.04897$.

\section{Methods}
\label{sec:meth}

In this section, we first demonstrate potential biases imposed on empirical relations by observational effects and discuss existing methods which partially account for them. Then we introduce the form of our model and discuss the measurements required to constrain its parameters. We follow this by detailing the T24 dataset and how it fulfills the requirements of our model. Finally, we provide the likelihood function used to fit our model to the T24 dataset. We also include numbered instructions for implementing our model with all optimal parameters under the ``Quick implementation'' subsection.

Throughout this section, we refer to the following galaxy properties: the emergent Ly$\alpha$ luminosity, $L_{\rm Ly\alpha}$, the velocity offset of the Ly$\alpha$ line with respect to the systemic redshift, $\Delta v$,  the H$\alpha$ luminosity, $L_{\rm H\alpha}$, and the UV continuum magnitude, ${\rm M}_{\rm UV}$. 

\begin{figure}
    \resizebox{\hsize}{!}
    {\includegraphics[width=0.4\textwidth]{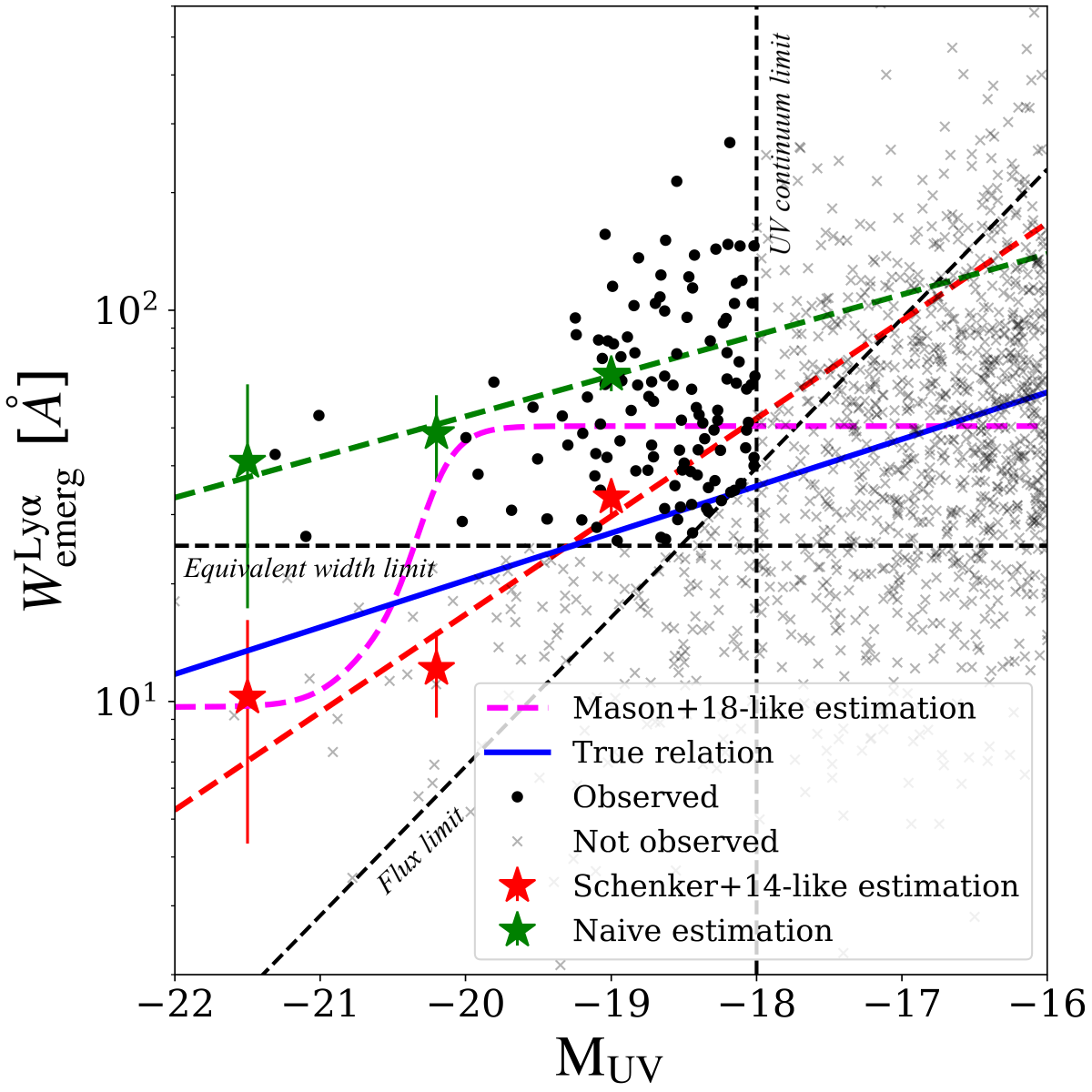}}
    \caption{Toy model illustrating the impact of observational selection bias on the inferred $\log_{10}W_{\rm emerg}^{\rm Ly\alpha}$--${\rm M}_{\rm UV}$ relation. The black dashed lines show the completeness limits of a realistic Ly$\alpha$ survey (taken from \citealt{debarros17}). The blue curve traces the mean of a fiducial normal $\log_{10}W_{\rm emerg}^{\rm Ly\alpha}$--${\rm M}_{\rm UV}$ relation. $1250$ samples drawn from this distribution are shown with black dots (observed) and grey x's (non-observed). The green stars and line show the relation one obtains by linear regression to the observed samples. The red stars and line show the relation one obtains by linear regression on the observed samples, corrected for $W_{\rm emerg}^{Ly\alpha}$ completeness using the method of \cite{schenker14}. The pink dashed curve shows the mean relation one obtains by the fitting procedure of \cite{mason2018}.}
    \label{fig:spur}
\end{figure}

\subsection{Observational effects}
\label{sec:obs}

An ideal survey of all  galaxy spectra in the Universe would provide a direct measurement of the joint probability distribution $P(L_{\rm Ly\alpha},\Delta v, L_{\rm H\alpha}, {\rm M}_{\rm UV})$. However, real surveys include various selection effects. In the simplest case, selection amounts to placing a conditional such that only a subset of an intrinsic distribution is observed\footnote{In the more generic case, observational effects may alter the observed data such that the observed distribution is a transformation of the intrinsic distribution (e.g., slit losses reducing line fluxes of more extended emission, c.f. \citealt{napolitano25}).
Here we assume that the treatment of {\it JWST} slit losses in the T24 analysis was correct, and therefore assume the simple case throughout this work.}. 

Defining the Ly$\alpha$/nebular properties of interest as:
\begin{equation}
\label{eq:xa}
    x_{\alpha}=\{\log_{10}L_{\rm Ly\alpha},\;\Delta v,\;\log_{10}L_{\rm H\alpha}\}
\end{equation}
 we would like to know the underlying (i.e. ``true") distribution $P(x_\alpha,{\rm M}_{\rm UV})$.  However, we can only observe the conditional distribution $P(x_\alpha,{\rm M}_{\rm UV}|{\rm obs})$, where ${\rm obs}$ is the condition that an object with properties $\{x_\alpha,{\rm M}_{\rm UV}\}$ is observed. These are related by the following:
\begin{equation}
    P(x_\alpha,{\rm M}_{\rm UV})=P(x_\alpha,{\rm M}_{\rm UV}|{\rm obs})P({\rm obs}),
\end{equation}
\noindent demonstrating that a measurement of $P(x_\alpha,{\rm M}_{\rm UV})$ requires an estimate of $P({\rm obs})$.

Figure \ref{fig:spur} illustrates the simple example of a set of galaxies with measured ${\rm M}_{\rm UV}$ and Ly$\alpha$ equivalent widths $W_{\rm emerg}^{\rm Ly\alpha}$. Throughout this work we define the emergent Ly$\alpha$ equivalent width as 
\begin{multline}
    W^{\rm Ly\alpha}_{\rm emerg}(L_{\rm Ly\alpha},{\rm M}_{\rm UV})=\left(\frac{1215.67\;\AA}{2.47\cdot10^{15}\;{\rm Hz}}\right)10^{-0.4(51.6-{\rm M}_{\rm UV})}\\
    \times L_{\rm Ly\alpha}\left(\frac{1215.67}{1500}\right)^{-\beta({\rm M}_{\rm UV})-2},
    \label{eq:ew_theory}
\end{multline}
\noindent where we adopt 
\begin{equation}
    \beta({\rm M}_{\rm UV})=-0.2({\rm M}_{\rm UV}+19.5)-2.05
    \label{eq:beta}
\end{equation}
\noindent from the fit to {\it Hubble} Space Telescope UV slope measurements made by \cite{bouwens2014}. This set of galaxies is subjected to three selection criteria representative of those present in a typical Ly$\alpha$ survey: a minimum Ly$\alpha$ flux, minimum Ly$\alpha$ equivalent width, and a maximum ${\rm M}_{\rm UV}$; in this example we adopt the criteria from \cite{debarros17}: ${\rm W}_{\rm emerg}^{\rm Ly\alpha}\geq25\;\AA$, Ly$\alpha$ flux$>2.2\cdot10^{-18}$ erg s$^{-1}$ cm$^{-2}$, and ${\rm M}_{\rm UV}\leq-18$. Each selection limit is shown with a black dashed line, and galaxies which meet all three criteria are shown in black dots while galaxies which do not are shown in grey x's. A fiducial true mean $W_{\rm emerg}^{\rm Ly\alpha}-{\rm M}_{\rm UV}$ relation is shown in a blue line, from which the log-normally distributed equivalent widths are drawn. If one were to naively treat the trend in the observed galaxies as representing the true trend, then the green line would be obtained. Clearly, an estimate of $P({\rm obs}|{\rm M}_{\rm UV})$ is required to correct this relation.

Most previous work on estimating mean $W_{\rm emerg}^{\rm Ly\alpha}-{\rm M}_{\rm UV}$ relations  used spectroscopic follow-up of a photometric dataset complete down to some ${\rm M}_{\rm UV}$. In these cases, the galaxies with detected ${\rm M}_{\rm UV}$ and non-detected Ly$\alpha$ can be used to estimate the $P({\rm obs}|{\rm M}_{\rm UV})$.  One could estimate a simple step-function form for $P({\rm obs}|{\rm M}_{\rm UV})$ by splitting the data into ${\rm M}_{\rm UV}$ bins and dividing the number of galaxies with observed Ly$\alpha$ emission by the total number of galaxies in the bin.
M18 followed this approach in their empirical fit of the distribution of Ly$\alpha$ equivalent widths to a set of LBGs with VLT FORS2 spectra. In their analysis, they split their data into three ${\rm M}_{\rm UV}$ bins, computed the mean equivalent width and observed fraction in each, and then fit each with a $\tanh$ function. In their model, the product of these functions may be taken as the mean $W_{\rm emerg}^{\rm Ly\alpha}-{\rm M}_{\rm UV}$ relation. \cite{schenker14} performed a similar exercise on a set of Lyman-break galaxies (LBGs) with MOSFIRE spectroscopic follow-up from the Keck telescope. However, instead of directly estimating $P({\rm obs}|{\rm M}_{\rm UV})$, \cite{schenker14} performed a linear regression on all galaxies, treating those with no detected Ly$\alpha$ emission as having an equivalent width of zero. In Figure \ref{fig:spur}, we show the fits one would obtain by following the approaches of \cite{schenker14} and M18 on our mock data. While these models provide a better approximation of $P({\rm W}_{\rm emerg}^{\rm Ly\alpha}|{\rm M}_{\rm UV})$ than the naive fit, they do not recover the true mean relation. Furthermore, they do not capture the intrinsic scatter around this mean relation.

In addition to observational selection effects, we assume that the infalling CGM attenuates all Lyman alpha flux blueward of the galaxy's circular velocity (e.g. \cite{barkana03}), computed as: 
\begin{equation}
    v_{\rm circ}=\left[10GM_hH(z)\right]^{1/3},
    \label{eq:vcirc}
\end{equation}
\noindent $G$ is the constant of universal gravitation, $M_h$ is the halo mass\footnote{Since our model does not include $M_h$, we adopt the broken linear relation of \cite{mason2015} to relate $M_h$ to ${\rm M}_{\rm UV}$: $\log_{10}M_h\;[M_\odot]=\gamma({\rm M}_{\rm UV} + 20.0 + 0.26z) + 11.75$, where $\gamma=-0.3$ for ${\rm M}_{\rm UV}\geq -20.0 - 0.26z$, and $\gamma=-0.7$ otherwise.}, and $H(z)$ is the Hubble constant at the redshift of the galaxy, $z$.  We 
assume that only galaxies with $\Delta v\geq v_{\rm circ}$ have observed Lyman alpha flux.

Summarizing observational effects which depend on survey properties as $\mathcal{T}$, we may say that the distribution of Ly$\alpha$ parameters measured in a real survey is $P(L_{\rm Ly\alpha},\Delta v, L_{\rm H\alpha}, {\rm M}_{\rm UV}|\mathcal{T},\Delta v>v_{\rm circ})$. In what follows, we recover $P(L_{\rm Ly\alpha},\Delta v, L_{\rm H\alpha}, {\rm M}_{\rm UV})$ by designing a model and calibration scheme which allow us to accurately estimate $P(\mathcal{T},\Delta v>v_{\rm circ})$.

\subsection{Empirical model}
\label{sec:model}

The relationship between the spectral properties of interest, $x_\alpha$ (c.f. Eq. \ref{eq:xa}), and any other galaxy properties, $x_{\rm gal}$, can generically be described by the joint probability distribution $P(x_{\alpha},x_{\rm gal})$. In the context of our model, we assume the only relevant functional dependence is the UV magnitude, $x_{\rm gal}=\{{\rm M}_{\rm UV}\}$.
We take advantage of the identity $P(x_{\alpha},{\rm M}_{\rm UV})=P(x_{\alpha}|{\rm M}_{\rm UV})P({\rm M}_{\rm UV})$ and the fact that $P({\rm M}_{\rm UV})$ is a known (observed) quantity, allowing us to solve only for the conditional distribution $P(x_{\alpha}|{\rm M}_{\rm UV})$.
Note that fitting a conditional PDF is computationally easier and can be done with a smaller sample, compared to fitting a joint PDF, as discussed further below.

 We fix $P({\rm M}_{\rm UV})$ to the Schechter fit of the $z\sim5$ {\it Hubble} UV luminosity function (UVLF) of \cite{bouwens21}:
\begin{multline}
    P({\rm M}_{\rm UV})\propto (0.4\ln10)\cdot0.79\cdot\left(10^{-0.4({\rm M}_{\rm UV}+21.1)}\right)^{-0.74}\\
    \times\exp\left\{-10^{-0.4({\rm M}_{\rm UV}+21.1)}\right\}.
    \label{eq:uvlf}
\end{multline}
  Importantly, the parameters in $x_\alpha$ are not independently distributed; for example, the H$\alpha$ luminosity of a galaxy typically increases with its Ly$\alpha$ luminosity since both trace the star formation rate (e.g., \citealt{kennicut98}). To account for these covariances, we model the conditional distribution $P(x_{\alpha}|{\rm M}_{\rm UV})$ as a multivariate Gaussian\footnote{Because a galaxy's luminosity can be expressed as a product of only weakly correlated terms (e.g. number of young stars, stellar SEDs, ISM/CGM opacity along a sightline), the central limit theorem could be used to roughly motivate log-normal luminosity distributions for a given galaxy population (e.g. at a fixed UV magnitude).} whose mean is a function of ${\rm M}_{\rm UV}$, i.e.,
\begin{equation}
\label{eq:gaussian}
    P(x_\alpha|{\rm M}_{\rm UV})\propto \exp\left\{-\frac{1}{2}(x_\alpha-\mu({\rm M}_{\rm UV}))^{\rm T}\Sigma^{-1}(x_\alpha-\mu({\rm M}_{\rm UV}))\right\},
\end{equation}
\noindent where the means scale linearly with ${\rm M}_{\rm UV}$ such that $\mu({\rm M}_{\rm UV})=m\cdot{\rm M}_{\rm UV}+b$, where $m$ and $b$ are 3-dimensional vectors, and $\Sigma$ is the symmetric 9-dimensional  covariance matrix of the distribution.
%

\begin{figure*}
    \resizebox{\hsize}{!}
    {\includegraphics[trim={0 80 0 120},clip,width=\textwidth]{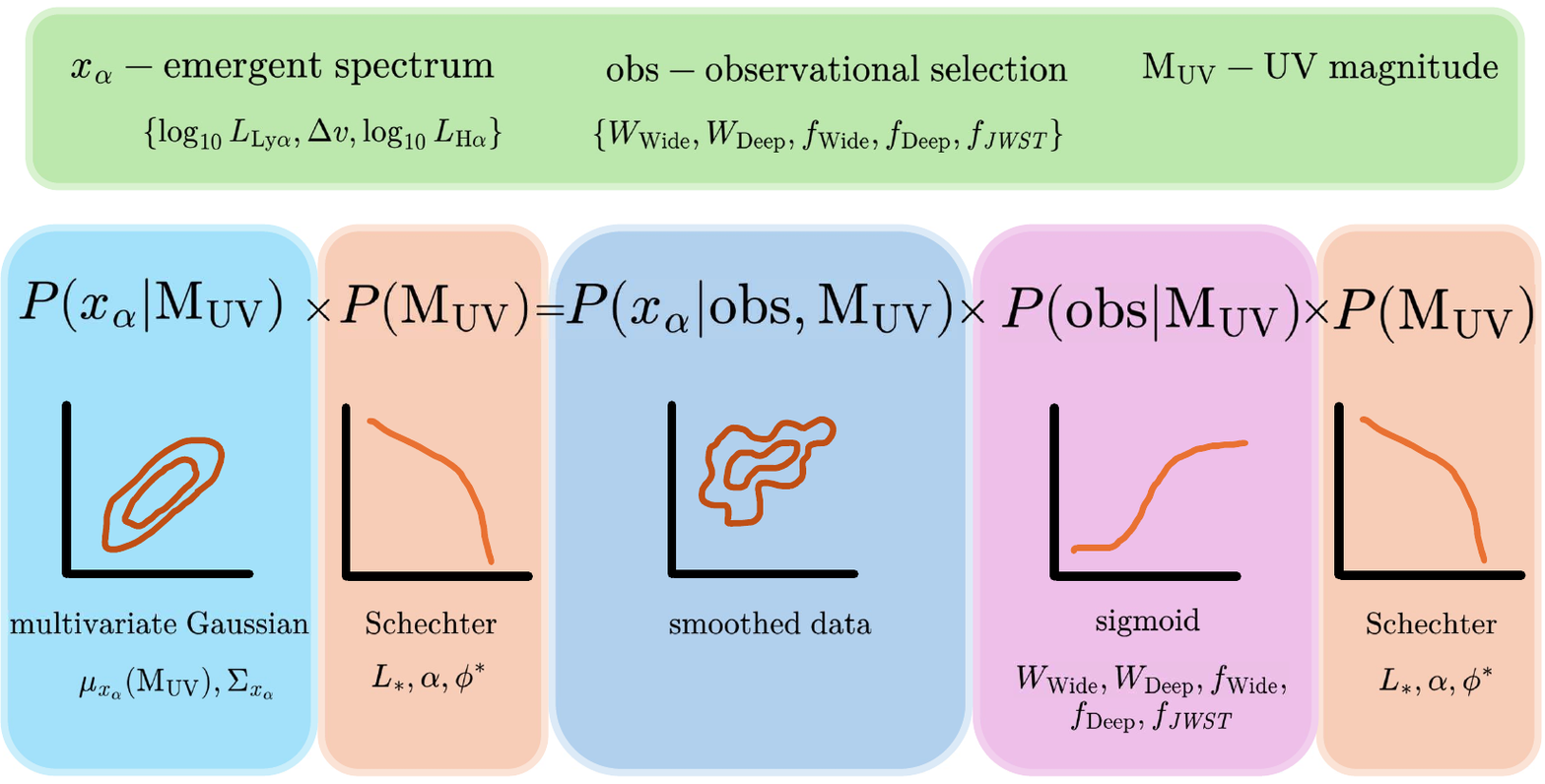}}
    \caption{Schematic overview of our model. We model $P(x_\alpha|{\rm M}_{\rm UV})$ as a multivariate Gaussian distribution conditioned on ${\rm M}_{\rm UV}$, and fit the parameters of this distribution, along with the parameters related to selection, labeled ${\rm obs}$, to the measured distribution of the data $P(x_\alpha|{\rm obs},{\rm M}_{\rm UV})$. We bin the data in ${\rm M}_{\rm UV}$ to take advantage of the Schechter fit to the measured UV luminosity function of \cite{bouwens21}. All observational selection functions are modeled as sigmoids, varying from $0$ to $1$, whose inflection points are free parameters.}
    \label{fig:flowchart}
\end{figure*}

\subsection{Observational sample}

\begin{table*}
    \centering
    \caption{Survey limiting fluxes reported by \cite{tang24}. Each flux corresponds to the $5\sigma$ detection limit of the relevant line in each survey.}
    \begin{tabular}{c|c|c}
        \hline
         Survey & Target line & Limiting flux [erg s$^{-1}$ cm$^{-2}$] \\
         \hline
         MUSE-Wide & Ly$\alpha$ & $2\cdot10^{-18}$\\ 
         MUSE-Deep (MOSAIC) & Ly$\alpha$ & $5\cdot10^{-19}$\\ 
         MUSE-Deep (UDF-10) & Ly$\alpha$ & $2.5\cdot10^{-19}$\\ 
         MUSE-Deep (MXDF) & Ly$\alpha$ & $1\cdot10^{-19}$\\ 
         \textit{JWST} FRESCO & H$\alpha$ & $2\cdot10^{-18}$\\ 
         \hline
    \end{tabular}
    \label{tab:flux_lim}
\end{table*}

\begin{table*}
    \centering
    \caption{Galaxy datasets with relevant limiting fluxes, UV continuum magnitudes, and Ly$\alpha$ equivalent widths.}
    \begin{tabular}{c|c|c|c|c}
        \hline
         Dataset & $W^{\rm Ly\alpha}_{\rm emerg,lim}$ [$\AA$] & $f_{\rm Ly\alpha,lim}$ [erg s$^{-1}$ cm$^{-2}$] & $f_{\rm H\alpha,lim}$ [erg s$^{-1}$ cm$^{-2}$] & ${\rm M}_{\rm UV,lim}$ \\
         \hline
         T24-Wide & $116$ & $1.8\cdot10^{-17}$ & $1.2\cdot10^{-18}$ & $-17.75$\\ 
         T24-Deep & $24$ & $2.7\cdot10^{-18}$ & $1.2\cdot10^{-18}$ & $-17.75$\\ 
         \hline
    \end{tabular}
    \label{tab:datasets}
\end{table*}

T24 presents a set of galaxies in the GOODS-S field at $z\sim5$ with detected Ly$\alpha$ and H$\alpha$ lines. These galaxies were observed with both VLT MUSE and {\it JWST} NIRCam, of which $24$ lie in the MUSE-Wide field and $36$ lie in the MUSE-Deep field (MUSE-Wide: \citealt{urrutia19}; MUSE-Deep: \citealt{bacon17,bacon23}; \textit{JWST} FRESCO: \citealt{oesch23}). The T24 sample includes only galaxies with ${\rm M}_{\rm UV}\gtrapprox-17.75$ and with line fluxes exceeding the $5\sigma$ flux limit of each survey. T24 reports the Ly$\alpha$ luminosity, H$\alpha$ luminosity, velocity offset of the Ly$\alpha$ line, and UV continuum magnitude of each galaxy in their sample. Table \ref{tab:flux_lim} lists the $5\sigma$ limiting flux for each survey. We divide the T24 sample into two datasets:

\begin{itemize}
    \item T24-Deep. All galaxies detected in Ly$\alpha$ by MUSE-Deep in the MOSAIC, UDF-10, and MXDF fields. Each galaxy also has an H$\alpha$ detection from \textit{JWST} FRESCO.
    \item T24-Wide. All galaxies detected in Ly$\alpha$ by MUSE-Wide, excluding those also detected by MUSE-Deep. Each galaxy also has an H$\alpha$ detection from \textit{JWST} FRESCO.
\end{itemize}

\noindent The observational limits of these datasets, summarized in Table \ref{tab:datasets}, were fit during the calibration of our model, where the effective equivalent width limits are imposed by the fixed ${\rm M}_{\rm UV}$ limit.

The T24 galaxy sample offers two distinct advantages: (i) the galaxies lie in a field with overlapping MUSE and JWST grism observations, providing measurements of both the Ly$\alpha$ line and H$\alpha$ line in all reported sources, and (ii) the exposure times are nearly uniform across the observing area, allowing us to characterize the completeness in terms of thresholds in Ly$\alpha$, H$\alpha$, and continuum emission. The former gives us a handle on the escape fraction of Ly$\alpha$ from galaxies, and the latter provides well-defined priors for the selection effects.

\subsection{Model calibration}

In this section, we describe how we fit the free parameters of the model described in Section \ref{sec:model} to the T24 datasets. In what follows, $x_{\alpha}$ refers to a spectral parameter vector sampled from our model (c.f. Eq. \ref{eq:xa}), while $\{x_{\alpha}^{\rm T24,Wide},\;x_{\alpha}^{\rm T24,Deep}\}$ refer to the parameters quoted in the T24-Wide and T24-Deep datasets, respectively. We solve for values of $m$, $b$, and $\Sigma$ which satisfy the condition:

\begin{multline}
    P(x_\alpha|{\rm M}_{\rm UV})P({\rm M}_{\rm UV})=\\P(x_\alpha|{\rm M}_{\rm UV},\mathcal{T}_{\rm T24,i})P(\mathcal{T}_{\rm T24,i}|{\rm M}_{\rm UV})P({\rm M}_{\rm UV}),
    \label{eq:calibration}
\end{multline}

\noindent where the indices $i$ correspond to the T24-Wide and -Deep datasets and the observational selection functions, $\mathcal{T}_{24,i}$, are sigmoid functions ranging from $0$ to $1$ with midpoints parameterized by their equivalent width completeness limits $W_{\rm Wide}$ and $W_{\rm Deep}$, and their flux limits $f_{\rm Wide}$, $f_{\rm Deep}$, and $f_{\it JWST}$. 
$P(x_\alpha|{\rm M}_{\rm UV},\mathcal{T}_{\rm T24,i})$ and $P({\rm M}_{\rm UV})$ correspond to the smoothed distribution of T24 data and the {\it Hubble} UVLF, respectively, so they remain fixed during our calibration. Meanwhile, the parameters of the {\it intrinsic} distribution of Ly$\alpha$ properties (i.e. before applying selection effects), $P(x_\alpha|{\rm M}_{\rm UV})$, and those of the observational selection functions, $P(\mathcal{T}_{\rm T24,i}|{\rm M}_{\rm UV})$, are free parameters fit during calibration. 
A schematic summarizing our approach is shown in Figure \ref{fig:flowchart}.

Our model assumes that $P(x_\alpha|{\rm M}_{\rm UV})$ is a multivariate Gaussian whose mean is a linear function of M$_{\rm UV}$. Therefore we have $15$ free parameters of the model ($m$ and $b$ are each $3$-dimensional, while $\Sigma$ is $9$-dimensional; c.f. Eq. \ref{eq:gaussian}), in addition to $5$ nuisance parameters related to selection (c.f. Table \ref{tab:datasets} and Appendix \ref{apdx:params}). Since performing regression in $20$ dimensions using only the T24 dataset would be challenging, we reduce the dimensionality of the problem by splitting it in two maximum likelihood estimation (MLE) steps:

\begin{enumerate}
    \item In the first step, we solve for the covariance matrix of the {\it intrinsic} distribution of Ly$\alpha$ parameters, which we treat as a weighted sum of four basis matrices: a $4$-dimensional MLE.
    \item In the second step, we solve for the $6$ mean parameters and $3$ variance parameters in the orthonormal basis, plus the $5$ nuisance parameters: a $14$-dimensional MLE.
\end{enumerate}

In both steps, we transformed our data to a space with zero mean and unit variance, defining $x_{\alpha}=\sigma_{\rm Ly\alpha}x_{\alpha,0}+\mu_{\rm Ly\alpha}$, where $x_{\alpha,0}$ is the unit variance zero mean parameter space. We used the mean and standard deviation from the T24-Deep dataset for this transformation since it contains more samples and should better reflect the tails of the distribution.

For the first step, we define the transformation $A$ and eigenvector $u$, where $x_{\alpha,0}=Au$. We solve for $A$ by minimizing the objective function
\begin{equation}
    {\rm obj}=\sum_i\left(u-x_{\alpha,i}^{T24,\rm Deep}\right)^{\rm T}{\rm Cov}^{-1}(u)\left(u-x_{\alpha,i}^{T24,\rm Deep}\right),
\end{equation}
\noindent describing the similarity of the samples $u\sim P(x_{\alpha}|\mathcal{T}_{\rm T24,Deep})$ to the data $x_{\alpha}^{T24,\rm Deep}$\footnote{We expect the correlations in the MUSE-Deep sample to at least partially reflect the intrinsic correlations between Ly$\alpha$ parameters. The MUSE-Wide selection cuts out far more objects, meaning that the correlations in the surviving sample of objects mostly reflects the correlations imposed by the selection itself.}, using differential evolution \citep{storn97}. We define $A$ as a sum of basis transformations, $A=c_1\mathcal{I}+c_2 S_1+c_3 S_2 + c_4 S_3$, where $\mathcal{I}$ is the identity matrix, and $S_i$ are the first, second, and third transformation matrices of the SO(3) special orthogonal Lie group (listed in Appendix \ref{apx:matrices}), and $c_i$ are the free parameters optimized by differential evolution.

In the second step we solve for the remaining free parameters of the multivariate Gaussian and the observational selection functions by matching both sides of Eq. \ref{eq:calibration}. Our likelihood function, again minimized by differential evolution \citep{storn97}, is a Gaussian distance computed on the following quantities: the observed fraction $f_{\rm obs}$, the emergent Ly$\alpha$ equivalent width $W_{\rm emerg}^{\rm Ly\alpha}$, the velocity offset $\Delta v$, and the Ly$\alpha$ escape fraction $f_{\rm esc}^{\rm Ly\alpha}$. The functional form of this likelihood is given in Appendix \ref{appdx:like2}. We define $f_{\rm obs}({\rm M}_{\rm UV})$ as the fraction of total galaxies at a given ${\rm M}_{\rm UV}$ in a given field of view which would be included in a dataset given its observational selection function:
\begin{equation}
    f_{\rm obs}({\rm M}_{\rm UV}|\mathcal{T})=\frac{P({\rm obs}|\mathcal{T},{\rm M}_{\rm UV})}{P({\rm M}_{\rm UV})}.
\end{equation}
\noindent In computing the escape fraction of Ly$\alpha$, we assume that the intrinsic Ly$\alpha$ luminosity arises from case A recombination \citep{osterbrock89,hayes10,matthee16,yang17,chen24,saxena24}, where $L_{\rm Ly\alpha}^{\rm intr}=11.4 L_{\rm H\alpha}$. Therefore, 
\begin{equation}
    f_{\rm esc}^{\rm Ly\alpha}(L_{\rm Ly\alpha},L_{\rm H\alpha})=\frac{L_{\rm Ly\alpha}}{11.4L_{\rm H\alpha}}.
    \label{eq:fesc_theory}
\end{equation}

\begin{figure}
    \resizebox{\hsize}{!}
    {\includegraphics[width=\textwidth]{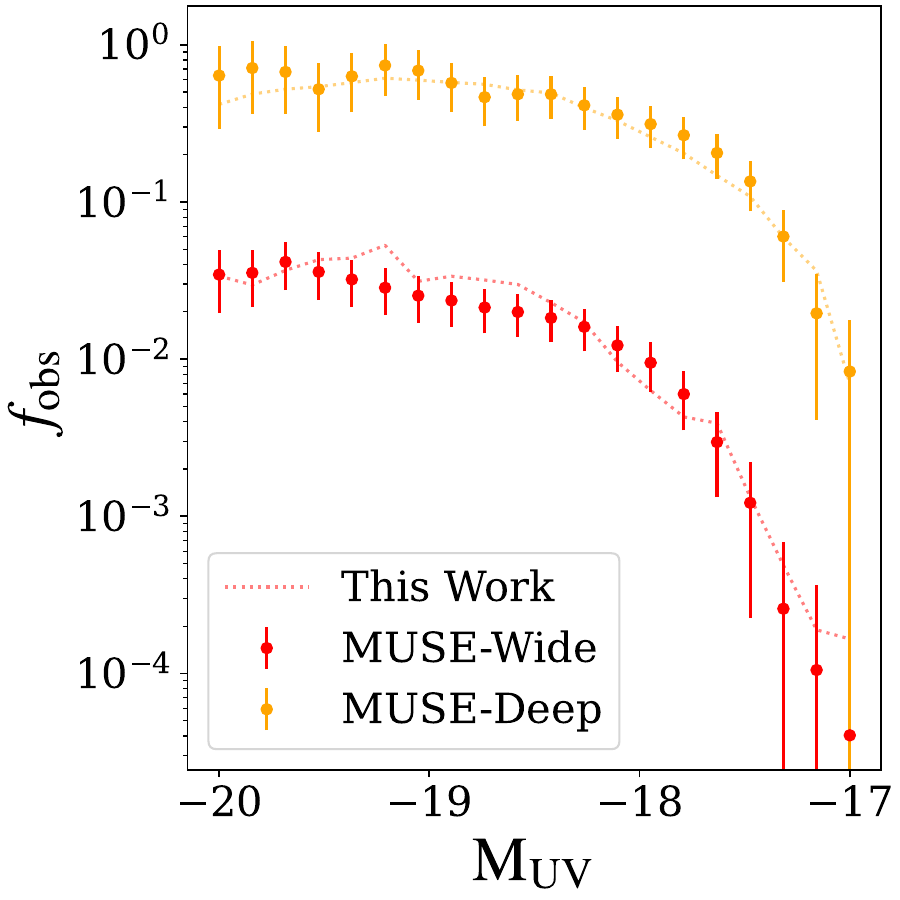}}
    \caption{The fraction of galaxies in the observation field observed in both Ly$\alpha$ and H$\alpha$ as a function of ${\rm M}_{\rm UV}$, shown in orange for T24-Deep and red for T24-Wide. Points with error bars are estimated from the T24 data, while dotted lines show the same statistic computed from a realization of our model.}
    \label{fig:fobs}
\end{figure}

Figure \ref{fig:fobs} shows the fraction of galaxies in the observation field observed in both Ly$\alpha$ and H$\alpha$ as a function of ${\rm M}_{\rm UV}$ for both the T24-Deep and T24-Wide samples, as well as those of our calibrated model. Our model correctly captures both the trend of decreasing $f_{\rm obs}$ with decreasing UV luminosity as well as the global difference in $f_{\rm obs}$ between the deep and wide fields. The observed fraction constrains the distance between the mean of $P(x_{\alpha})$ and the observed mean, $P(x_{\alpha}|\mathcal{T})$. Measuring the observed mean for two different observational selection functions allows the model to estimate the variance in $P(L_{\rm Ly\alpha},L_{\rm H\alpha}|{\rm M}_{\rm UV})$.

\subsection{Quick implementation}
\label{sec:quick}

We obtain the following procedure for sampling galaxy Ly$\alpha$ properties:

\begin{enumerate}
    \item Draw ${\rm M}_{\rm UV}\sim P({\rm M}_{\rm UV})$
    \item Draw $u\sim P(u|{\rm M}_{\rm UV})=\mathcal{N}(\mu({\rm M}_{\rm UV}), \sigma)$
    \item Transform $x_{\alpha,0}=Au$
    \item Transform $x_{\alpha}=\sigma_{\rm Ly\alpha}x_{\alpha,0}+\mu_{\rm Ly\alpha}$.
\end{enumerate}

\noindent Here, $P({\rm M}_{\rm UV})$ is the UVLF in Eq. \ref{eq:uvlf}, the eigenvectors $u$ follow the distributions

\begin{equation}
    \begin{split}
        & u_1\sim\mathcal{N}\left(8.7\cdot10^{-2}{\rm M}_{18.5}-0.51,\;0.7\right) \\
        &u_2\sim\mathcal{N}\left(-0.57{\rm M}_{18.5}-0.85\;0.49\right) \\
        &u_3\sim\mathcal{N}\left(-0.38{\rm M}_{18.5}-0.31,\;0.26\right),
    \end{split}
\end{equation}

\noindent where ${\rm M}_{18.5}\equiv{\rm M}_{\rm UV}+18.5$, the change of basis transformation is 
\begin{equation}
    A=
    \begin{bmatrix}
        1 & 1 & 1/3 \\
        -1 & 1 & -1 \\
        -1/3 & 1 & 1 \\
    \end{bmatrix},
\end{equation}
\noindent and the mean and standard deviation of the data distribution are $\mu_{\rm Ly\alpha}=\left\{42.47,\;200.18,\;42.03\right\}$ and $\sigma_{\rm Ly\alpha}=\left\{0.42,\;99.7,\;0.39\right\}$, respectively. Appendix \ref{apdx:params} includes a table of all model parameters. We also include a more concise representation of the model as a multivariate normal distribution in Appendix \ref{apdx:mult_norm}. The code used for calibration and sampling from our model is publicly available\footnote{\url{https://github.com/samgagnon/empirical_lya}}.

\section{Results}
\label{sec:res}

\begin{figure*}
    \resizebox{\hsize}{!}
    {\includegraphics[width=\textwidth]{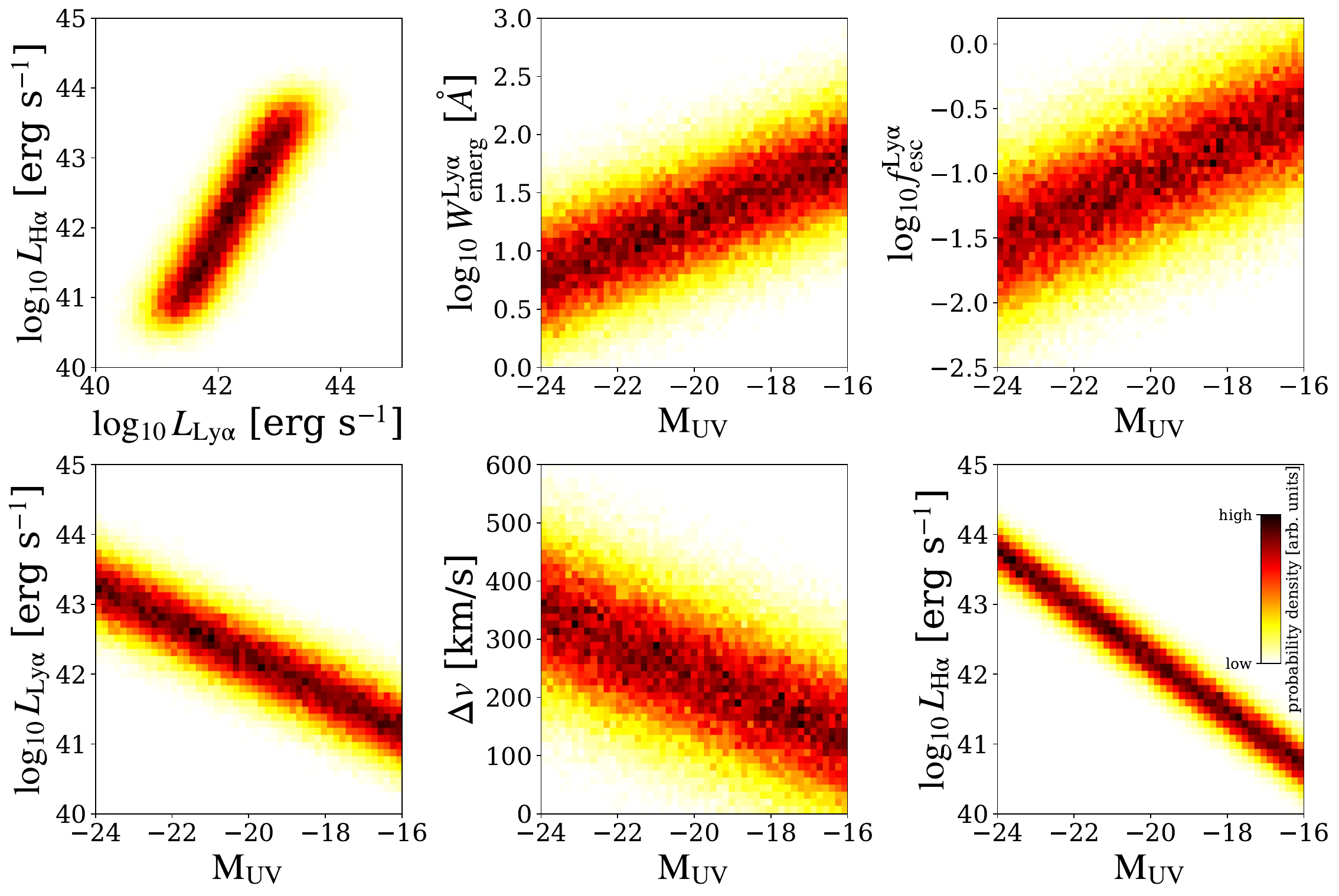}}
    \caption{Scalings between select Ly$\alpha$ parameters in our model. For better visualization of the trends at the rare/bright end, this figure was generated by uniformly sampling ${\rm M}_{\rm UV}$ over the range shown. The Ly$\alpha$ luminosity, H$\alpha$ luminosity, and velocity offset of the Ly$\alpha$ line all increase with increasing UV continuum luminosity (bottom row) whereas the Ly$\alpha$ emergent equivalent width and escape fraction increase for decreasing UV continuum luminosity (top row).}
    \label{fig:var}
\end{figure*}

In this section, we first illustrate the parameter covariances in our model and then show that our model can reproduce the distribution of the T24 data. We list closed functional forms for the marginal distributions of each Ly$\alpha$ parameter conditioned on ${\rm M}_{\rm UV}$. We also provide functional forms for the marginal distributions of $W_{\rm Ly\alpha}^{\rm emerg}({\rm M}_{\rm UV})$ and $f_{\rm esc}^{\rm Ly\alpha}({\rm M}_{\rm UV})$.

Figure \ref{fig:var} illustrates the scalings between Ly$\alpha$ parameters in our model with various 2D slices of our full parameter distribution. In each slice, we sample ${\rm M}_{\rm UV}\sim{\rm Unif}[-24,-16]$ for clarity. The bottom row shows how our Ly$\alpha$ parameters scale with ${\rm M}_{\rm UV}$. Each marginal distribution, $P(x_\alpha|{\rm M}_{\rm UV})$, is normally-distributed with a mean that varies linearly with UV magnitude and a constant standard deviation: 
\begin{equation}
    \log_{10}\left(\frac{L_{\rm Ly\alpha}({\rm M}_{\rm UV})}{{\rm\;erg\;s^{-1}}}\right)=-0.28({\rm M}_{\rm UV}+18.5)+41.86\pm0.36,
    \label{eq:lya}
\end{equation}
\begin{equation}
    \Delta v({\rm M}_{\rm UV})=-27.92({\rm M}_{\rm UV}+18.5)+197.19\pm89.17{\rm\;km\;s^{-1}},
    \label{eq:dv}
\end{equation}
\begin{equation}
    \log_{10}\left(\frac{L_{\rm H\alpha}({\rm M}_{\rm UV})}{{\rm\;erg\;s^{-1}}}\right)=-0.38({\rm M}_{\rm UV}+18.5)+41.64\pm0.24,
    \label{eq:ha}
\end{equation}
\noindent each written as mean $\pm$ standard deviation. The mean emergent Ly$\alpha$ and H$\alpha$ luminosities increase with increasing UV luminosity, as expected since all scale with the star formation rate of a galaxy. The velocity offset of the emergent Lyman alpha line with respect to its systemic redshift, $\Delta v$, also increases with UV luminosity, albeit with high scatter.

\begin{figure*}
    \resizebox{\hsize}{!}
    {\includegraphics[width=\textwidth]{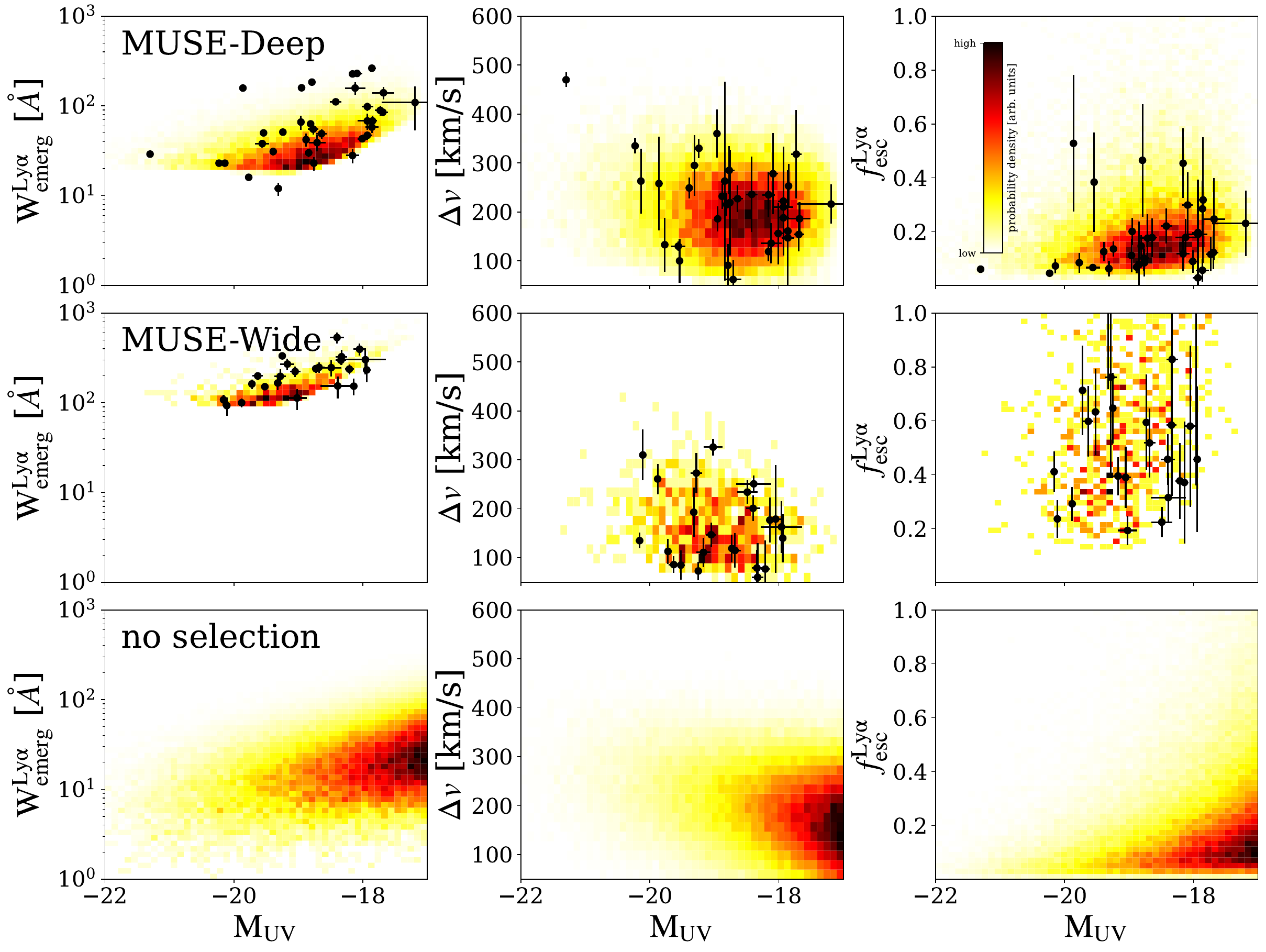}}
    \caption{The distribution of Ly$\alpha$ properties predicted by our model (red-orange probability density function) and measured by \cite{tang24} (black points). The top row applies the flux completeness limits of MUSE-Deep and the {\it JWST} FRESCO survey, the second row applies that of MUSE-Wide and the {\it JWST} FRESCO survey, and the third row applies no selection effects. The PDF in each panel is normalized independently for visibility.}
    \label{fig:hist}
\end{figure*}

Now we investigate the joint distribution $P(x_{\rm Ly\alpha}, {\rm M}_{\rm UV})$ produced by our model, which we produce by repeating the above exercise while using the UV luminosity function (Eq. \ref{eq:uvlf}) for $P({\rm M}_{\rm UV})$. 
Figure \ref{fig:hist} shows the joint distributions of $W_{\rm emerg}^{\rm Ly\alpha}$, $\Delta v$, and $f_{\rm esc}^{\rm Ly\alpha}$ with ${\rm M}_{\rm UV}$ measured by \cite{tang24} and predicted by our model. The bottom row shows the intrinsic distribution of parameters, with no selection effects applied. The top and middle rows respectively include the selection effects of T24-Deep and T24-Wide (see Table \ref{tab:datasets} for selection criteria). Our model fits both the T24-Wide and T24-Deep datasets very well.

Combining Eq.s \ref{eq:lya}-\ref{eq:ha} with Eq.s \ref{eq:ew_theory} and \ref{eq:fesc_theory} yields $f_{\rm esc}^{\rm Ly\alpha}({\rm M}_{\rm UV})$ and $W_{\rm emerg}^{\rm Ly\alpha}({\rm M}_{\rm UV})$:
\begin{equation}
    \log_{10}f_{\rm esc}^{\rm Ly\alpha}({\rm M}_{\rm UV})=0.12({\rm M}_{\rm UV}+18.5)+0.28\pm0.46-\log_{10}\alpha_{\rm rec},
    \label{eq:fesc}
\end{equation}
\noindent where $\alpha_{\rm rec}$ is the ratio between the intrinsic luminosity of Ly$\alpha$ and that of H$\alpha$ ($\alpha_{\rm rec}=11.4$ for case A recombination and $\alpha_{\rm rec}=8.2$ for case B; \cite{baker38,osterbrock89,dijkstra14}), and 
\begin{equation}
    \log_{10}W_{\rm emerg}^{\rm Ly\alpha}({\rm M}_{\rm UV}){\rm\;[\AA]}=0.12({\rm M}_{\rm UV}+18.5)+1.49\pm0.36.
    \label{eq:ew_fit}
\end{equation}

\section{Discussion}
\label{sec:disc}

\begin{figure}
    \resizebox{\hsize}{!}
    {\includegraphics[width=\textwidth]{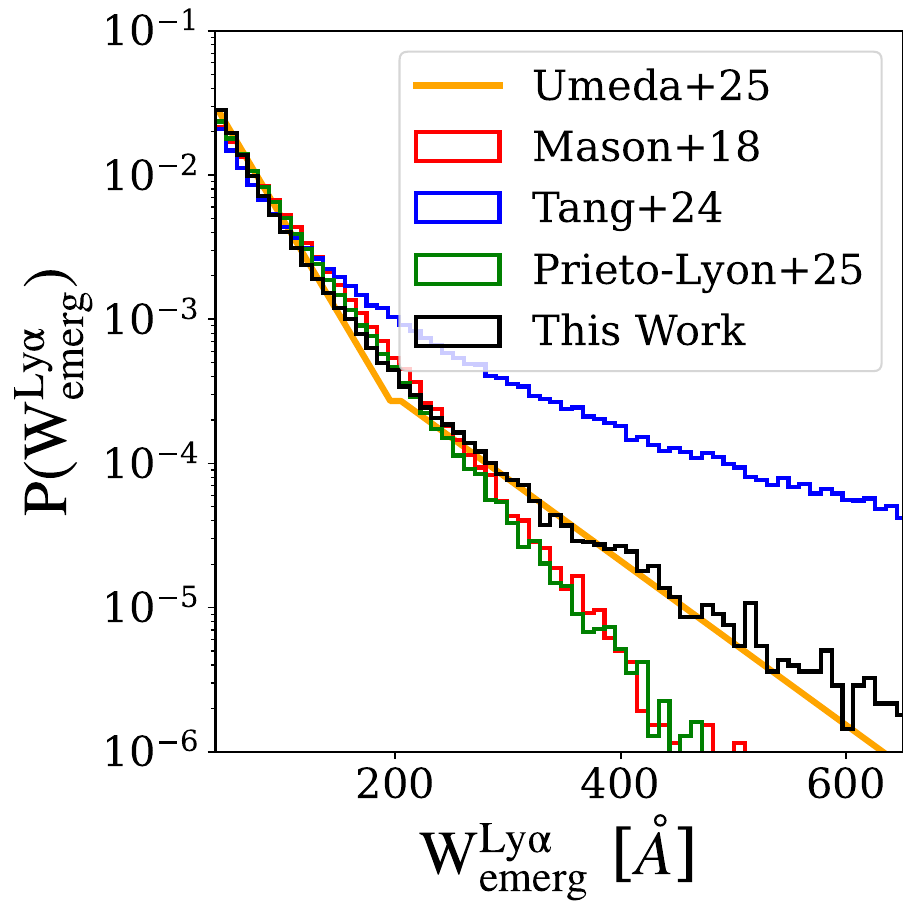}}
    \caption{The distribution of equivalent widths for $z\sim5$ galaxies with ${\rm M}_{\rm UV}<-18$ reported by \cite{umeda25} (corresponding to a double power law fit to the data) is shown with the yellow curve.  The corresponding marginal PDF predicted by our model is shown with the black histogram.  The agreement of the yellow curves and the black histogram is an important sanity check for our model, as it was calibrated to independent data, with very different observational selections. For comparison, we also show the EW PDFs predicted by \cite{mason2018}, \cite{tang24}, and \cite{prietolyon25}.  Our model matches the high EW tail of the PDF observed by Subaru better than previous estimates.}
    \label{fig:ewpdf}
\end{figure}

\begin{figure}
    \resizebox{\hsize}{!}
    {\includegraphics[width=\textwidth]{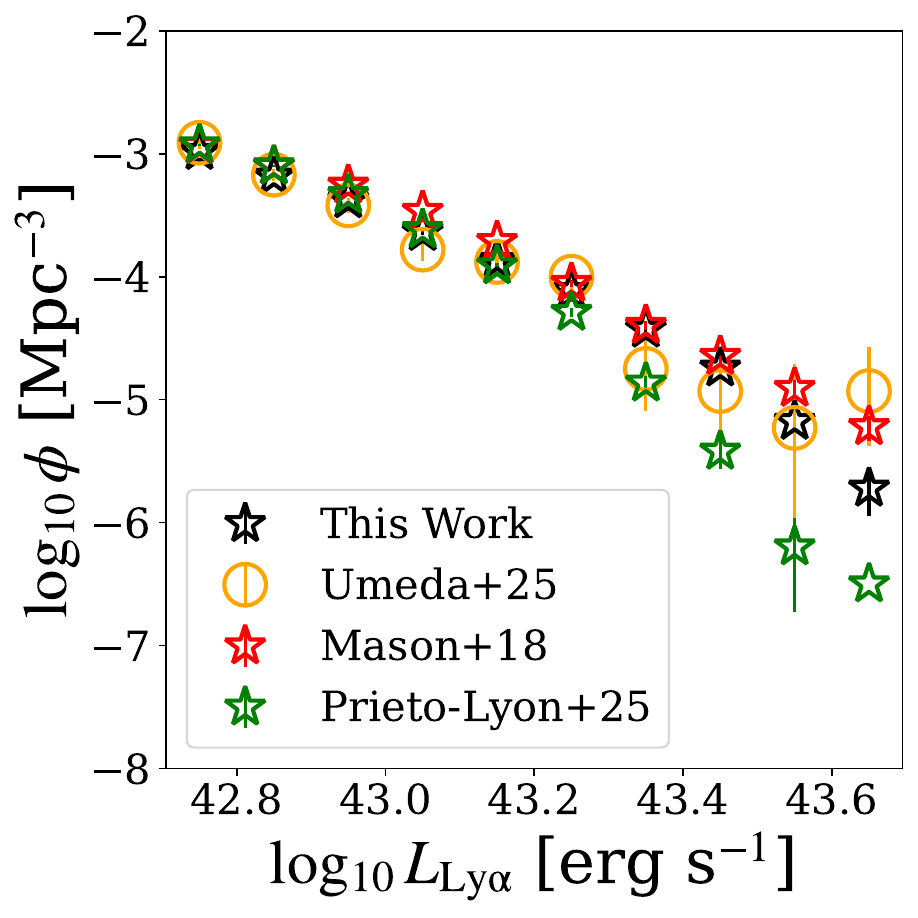}}
    \caption{The $z\sim5$ Ly$\alpha$ luminosity function measured by \cite{umeda25} and predicted by the models of \cite{mason2018}, \cite{prietolyon25}, and this work.}
    \label{fig:laelf}
\end{figure}

Here we compare our derived Ly$\alpha$ properties to previous works and discuss some implied trends.  

\subsection{Comparison with previous work}

Before examining physical implications and application of our model, we compare the statistics predicted by our model with those reported by \cite{mason2018}, 
\cite{prietolyon25}, \cite{tang24}, and \cite{umeda25}.\footnote{We opt not to compare our model with that of \cite{khoraminezhad25} since theirs is calibrated to objects at $z\sim2-3$, while ours focuses on $z\sim 5$.}

\cite{umeda25} reported a huge suite of $\sim20,000$ LAEs detected through Subaru photometry at $z\in[2.2,\;7.3]$ and computed several summary statistics of their LAE population, including the equivalent with distribution (EW PDF). Figures \ref{fig:ewpdf} and \ref{fig:laelf} show a double power law fit to the EW PDF and the Ly$\alpha$ luminosity function (LAELF) reported by \cite{umeda25} at $z\sim5$, together with those produced by realizations of our model and the models of \cite{mason2018} and \cite{prietolyon25}. We also show the EW PDF estimated by \cite{tang24}. These EW PDFs are marginalized over ${\rm M}_{\rm UV}\leq-18$.

\cite{tang24} estimated the EW PDF by fitting log-normal distributions to subsets of their data. We produced the blue curve shown in Figure \ref{fig:ewpdf} by taking a weighted average of their log-normal fits to Lyman-break-selected galaxies in the T24 data centered at ${\rm M}_{\rm UV}=-18.5$ and ${\rm M}_{\rm UV}=-19.5$, where we set the weights equal to the value of the UVLF at those magnitudes. The T24 model is fit to primarily low-$W_{\rm emerg}^{\rm Ly\alpha}$ objects using a log-normal distribution, so its accuracy decreases at the high $W_{\rm emerg}^{\rm Ly\alpha}$ relevant to the LAELF in Figure \ref{fig:ewpdf}.


\cite{mason2018} (M18) introduced an empirical model of Ly$\alpha$ emission which has since been applied in several works (e.g., \citealt{umeda25, lu25, gagnonhartman25, nikolic25}) and has been shown to reproduce the LAE LF at $z\sim2-6$ \citep{morales21}. The M18 model assumes that all galaxies have a probability $A({\rm M}_{\rm UV})$ of emitting in Ly$\alpha$ an equivalent width drawn from an exponential distribution of mean $W_c({\rm M}_{\rm UV})$ \citep{dijkstra12}. The line profile is assumed to be a Gaussian whose FWHM equals its $\Delta v$, and they assume that a neutral CGM absorbs all Ly$\alpha$ emission with $\Delta v<v_{\rm circ}$ (see Eq. \ref{eq:vcirc}). M18 modeled $A({\rm M}_{\rm UV})$ and $W_c({\rm M}_{\rm UV})$ as $\tanh$ functions fit to a sample of $z\sim6$ LAEs detected in the VANDELS ESO spectroscopic survey, a Large Program of the VLT \citep{pentericci18}. The procedure used by M18 to recover the $W_{\rm Ly\alpha}^{\rm emerg}-{\rm M}_{\rm UV}$ relation is prone to including unquantified bias (c.f. Section \ref{sec:obs} and Figure \ref{fig:spur}). We provide recalibrated versions of their functions which better match the measured EW PDF of \citealt{umeda25} in Appendix \ref{apdx:m18}.

When compared to the measured EW PDF from Subaru, M18 tends to overproduce low-$W_{\rm emerg}^{\rm Ly\alpha}$ and underproduce high-$W_{\rm emerg}^{\rm Ly\alpha}$ LAEs, especially beyond $W_{\rm emerg}^{\rm Ly\alpha}\sim 300$. In both regimes our model comes closer to the measured EW PDF. 
We attribute the accuracy of our model to its accurate determination of the variance in the ${\rm M}_{\rm UV}-{\rm W}_{\rm emerg}^{\rm Ly\alpha}$ relation (e.g., \citealt{dijkstra12}). This is made possible by: (i) having more flexible conditional PDFs; and (ii) the inclusion of both a deep and wide  field in our calibration dataset. We present further evidence of the usefulness of (ii) in Appendix \ref{apdx:ewpdf}. We note that our model diverges from the measured EW PDF at $W_{\rm Ly\alpha}^{\rm emerg}\approx200$ $\AA$: this is expected since \cite{umeda25} reports a double power law fit to the measured EW PDF, and we expect the transition between these domains to be smooth, likely closer to our model than the yellow curve. We stress that our model was exclusively calibrated on the T24 data, and therefore its ability to reproduce the EW PDF and LAE LF of \cite{umeda25} indicate the general applicability of the scaling relations and covariances reported in our study.

\cite{prietolyon25} explored the dependence of  $W_{\rm emerg}^{\rm Ly\alpha}$, $f_{\rm esc}^{\rm Ly\alpha}$, and $\Delta v$ on the UV magnitude and slope.  They used NIRCAM/grism spectra of galaxies in the GOODS-N field complete in ${\rm M}_{\rm UV}$ down to $-19$ \citep{eisenstein23}. The scalings found for $W_{\rm emerg}^{\rm Ly\alpha}$ and trend for$\Delta v$ qualitatively agree with our own, as does their reported lack of a significant correlation between $f_{\rm esc}^{\rm Ly\alpha}$ and ${\rm M}_{\rm UV}$, since this correlation is low for the UV-bright objects in their dataset. While the EW PDF of their model is consistent with that of M18, we find that their LAE LF underproduces Ly$\alpha$-bright objects relative to the \cite{umeda25} measurement (c.f. Figs. \ref{fig:ewpdf}-\ref{fig:laelf}).

\subsection{GS-z13 -- is it surprising to detect Ly$\alpha$ at $z\sim13$?}

\begin{figure}
    \resizebox{\hsize}{!}
    {\includegraphics[width=\textwidth]{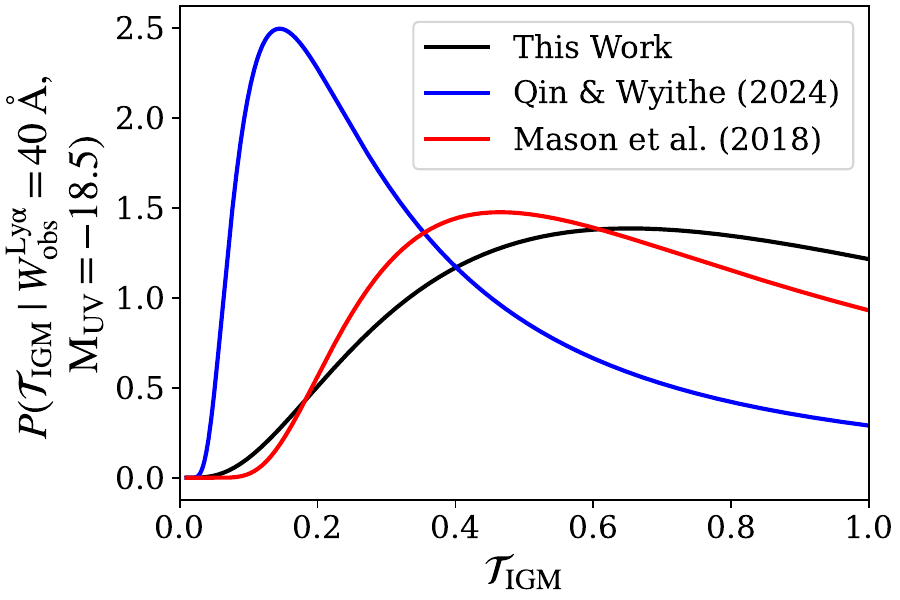}}
    \caption{Probability density functions of the IGM transmission fraction, $\mathcal{T}_{\rm IGM}$ in our model (black) and those of M18 (red) and \citealt{qin_wyithe24} (blue), conditioned on having an observed Ly$\alpha$ equivalent width of $40$ $\AA$, an emergent Ly$\alpha$ equivalent width larger than $40$ $\AA$, and ${\rm M}_{\rm UV}=-18.5$, consistent with the properties of GS-z13 \citep{witstok25}.}
    \label{fig:tigm_gsz13}
\end{figure}

\begin{figure}
    \resizebox{\hsize}{!}
    {\includegraphics[width=\textwidth]{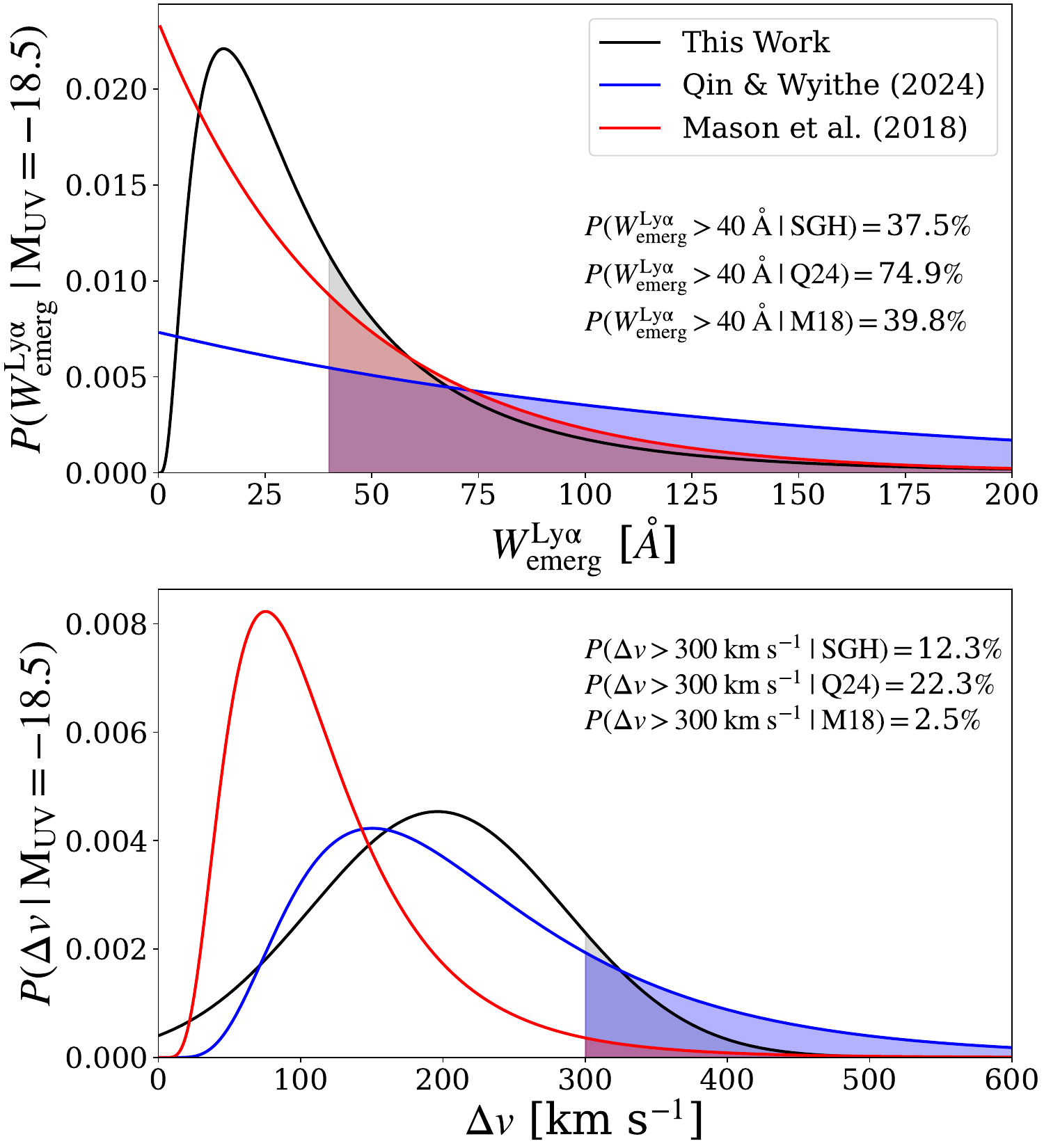}}
    \caption{The distributions of emergent Ly$\alpha$ equivalent widths (top) and red wing velocity offsets (bottom) for galaxies with ${\rm M}_{\rm UV}=-18.5$ in the empirical model presented in this work (black), as well as those presented by \cite{mason2018} (blue) and \cite{qin_wyithe24} (red). The figure also lists the cumulative probabilities of $W_{\rm emerg}^{\rm Ly\alpha}>40$ $\AA$ (top) and $\Delta v>300$ km s$^{-1}$ (bottom) in each model.}
    \label{fig:dv_gsz13}
\end{figure}

\begin{table}
    \centering
    \caption{Observed properties of GS-z13.}
    \begin{tabular}{c|c|c}
    \hline
    \textbf{Property} & \textbf{Units} & \textbf{Value} \\
    \hline
    ${\rm M}_{\rm UV}$ & - & $-18.5$ \\
    $W_{\rm obs}^{\rm Ly\alpha}$ & $\AA$ & $40.0\pm13.8$  \\
    Ly$\alpha$ flux & $10^{-19}$ erg s$^{-1}$ cm$^{-2}$ & $7.46\pm1.16$ \\
    H$\beta$ flux & $10^{-19}$ erg s$^{-1}$ cm$^{-2}$ & $<6.7$ \\
    \end{tabular}
    \label{tab:gsz13}
\end{table}

\cite{witstok25} presented GS-z13, a spectroscopically-confirmed Ly$\alpha$-emitting galaxy at $z\sim13$, whose relevant properties we list in Table \ref{tab:gsz13}. The detection of Ly$\alpha$ emission  in the early stages of the EoR (e.g. \citealt{qin24}) suggests either that GS-z13 is emitting significant Ly$\alpha$ at a high velocity offset relative to the neutral gas surrounding it (i.e. with a large $L_{\rm Ly\alpha}$ and $\Delta v$, see \citealt{dijkstra11}), and/or that it is sitting in an ionized bubble large enough to allow for significant Ly$\alpha$ transmission. 

Given these uncertainties, the {\it observed} equivalent width of $40$ $\AA$ can be treated as a lower limit to the {\it emergent} equivalent width.  The two quantities are related by a factor that encodes the wavelength-averaged transmission through the patchy EoR:
%
\begin{equation}
\label{eq:transmission}
    \mathcal{T}_{\rm IGM} = \frac{W_{\rm obs}^{\rm Ly\alpha}}{W_{\rm emerg}^{\rm Ly\alpha}} ~,
\end{equation}
\noindent where $\mathcal{T}_{\rm IGM}$ is close to unity post EoR (as is assumed in our model calibrated at $z\sim5$), and ranges to zero depending on the EoR stage and local morphology around the galaxy (e.g. \citealt{mesinger08,tingyi24}).
Evaluating Eqs. \ref{eq:ew_fit} at GS-z13's UV magnitude, $M_{\rm UV} = -18.5$, we obtain $W_{\rm emerg}^{\rm Ly\alpha}=30.49^{+39.36}_{-17.18}$ $\AA$ and $\Delta v=197.19\pm 89.17$ km s$^{-1}$, placing the object's observed equivalent width within $1\sigma$ of the mean of our model's distribution for the emergent equivalent width. From Eq. \ref{eq:transmission}, we obtain that the implied IGM transmission is $\mathcal{T}_{\rm IGM}=0.62^{+0.25}_{-0.26}$, where the upper and lower bounds represent $16$th and $84$th percentiles. Figure \ref{fig:tigm_gsz13} shows $P(\mathcal{T}_{\rm IGM}~|~W_{\rm obs}^{\rm Ly\alpha}=40\;\AA, W_{\rm emerg}^{\rm Ly\alpha}>40\;\AA, {\rm M}_{\rm UV}=-18.5)$ for our model as well as those of M18 and \citealt{qin_wyithe24}, who recalibrated the M18 relations to better match the properties of recently detected Ly$\alpha$-emitting galaxies from $2<z<11$ \citep{tang24,witstock24b}.

As discussed above, interpreting this value of the wavelength-averaged IGM transmission requires making assumptions about the local EoR morphology as well as the intrinsic velocity offset of the emergent Lyman alpha line.  We defer a detailed study to future work; however, we note that
\citet{mason_gronke20} estimated a minimum value of $\Delta v>300$ km s$^{-1}$ would be required to detect Lyman alpha at those redshifts/EoR stages.
Referring to Eq. \ref{eq:dv}, this minimum inferred value is only $\sim1.1 \sigma$ away from the mean of our distribution.

To put these in better context with previous studies, in Figure \ref{fig:dv_gsz13} we show our distributions of $W_{\rm emerg}^{\rm Ly\alpha}$ and $\Delta v$, evaluated at ${\rm M}_{\rm UV}=-18.5$ (black curves).  We also show the corresponding PDFs for the empirical models of M18 (red curves) and \cite{qin_wyithe24} (blue curves). 
We highlight the probability densities beyond $40$ $\AA$ and $300$ km s$^{-1}$, expected to correspond to the emergent values of GS-z13 as discussed above. All three models can accommodate the lower limit of $W_{\rm emerg}^{\rm Ly\alpha}\approx40$ $\AA$ with reasonable probabilities.
However, the relatively high value of $\Delta v$ derived by \citet{mason_gronke20} is considerably more likely in our model of the emergent profile and that of \cite{qin_wyithe24}, compared to the M18 model.


As our model describes Ly$\alpha$ properties with a multivariate normal distribution with a non-diagonal covariance, it also quantifies the probability of galaxies having a certain Ly$\alpha$ property conditioned on its other Ly$\alpha$ properties. For example, given a known Ly$\alpha$ equivalent width and ${\rm M}_{\rm UV}$, we may produce the conditional distribution of velocity offsets, $P(\Delta v|W_{\rm emerg}^{\rm Ly\alpha},{\rm M}_{\rm UV})$. In our model, the $\Delta v$ is inversely correlated with $W_{\rm emerg}^{\rm Ly\alpha}$, bringing $P(W_{\rm emerg}^{\rm Ly\alpha}>40\;\AA)$ down to $\sim13\%$ for $\Delta v=300$ km s$^{-1}$, and lower beyond that. The formulas we employed to produce conditional distributions of our model are provided in Appendix \ref{appdx:conditionals}.

\subsection{Outlier detection: a GN-z11 analog at $z\approx5$?}


\begin{table*}
    \centering
    \caption{The properties of GN-z11 \citep{bunker23} and MUSE-1670 \citep{tang24} compared with the mean values of the conditional distribution $P(W_{\rm emerg}^{\rm Ly\alpha},\Delta v,f_{\rm esc}^{\rm Ly\alpha}|{\rm M}_{\rm UV}=-21.5)$ predicted by our model, with $1\sigma$ bounds.}
    \begin{tabular}{c|c|c|c|c}
    \hline
    \textbf{ID} & ${\rm M}_{\rm UV}$ & $W_{\rm emerg}^{\rm Ly\alpha}$ [$\AA$] & $\Delta v$ [km s$^{-1}$] & $f_{\rm esc}^{\rm Ly\alpha}$\\
    \hline
    GN-z11 & $-21.5$ & $18.0$ & $555$ & $0.03\pm0.05$\\
    MUSE-1670 & $-21.3\pm0.01$ & $29.0\pm1.0$ & $470.0\pm15.0$ & $0.06\pm0.01$ \\
    \textit{This Work} & $-21.5$ & $13.49^{+17.41}_{-7.60}$ & $280.95\pm89.17$ & $0.10^{+0.20}_{-0.07}$ \\
    \hline
    \end{tabular}
    \label{tab:analog}
\end{table*}

\begin{figure}
    \resizebox{\hsize}{!}
    {\includegraphics[width=\textwidth]{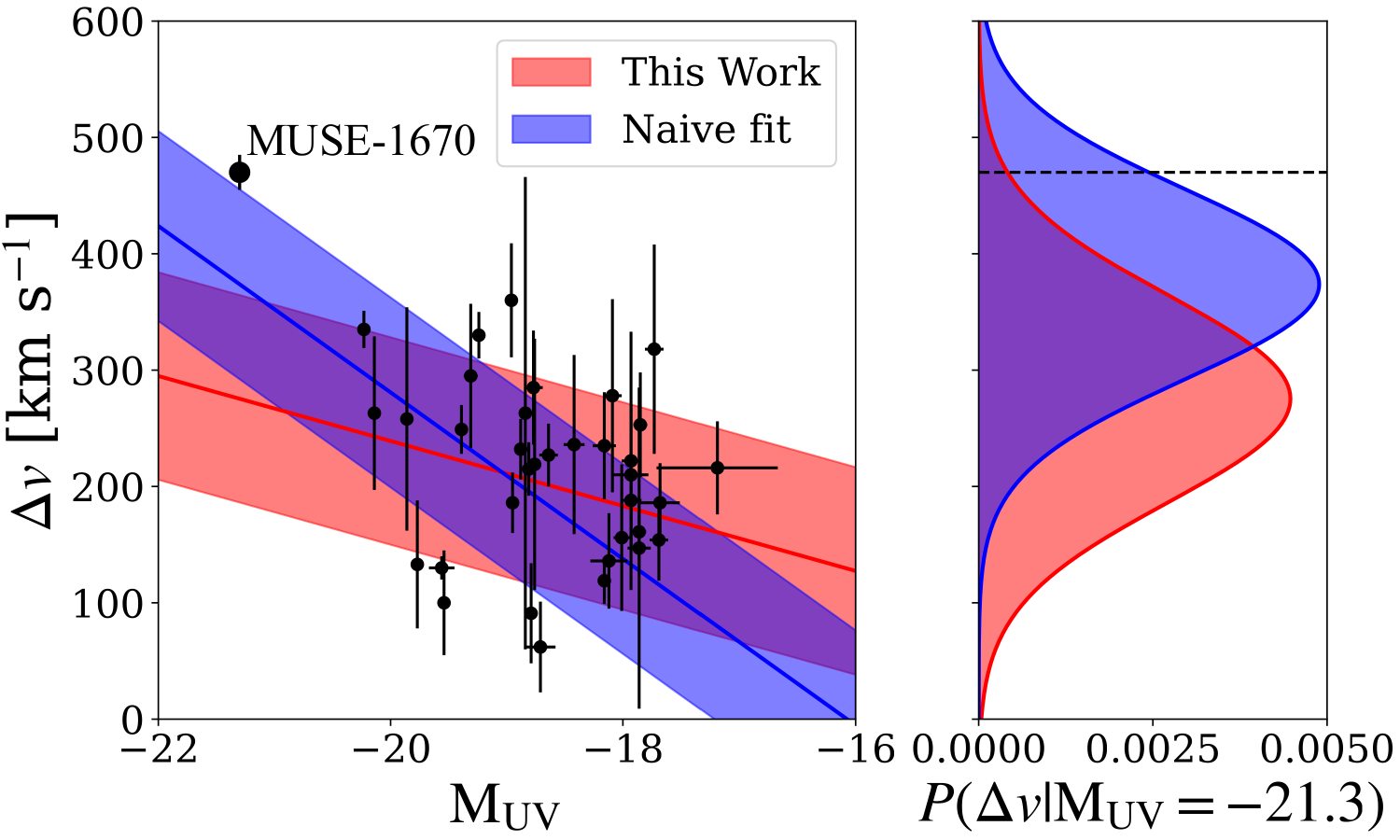}}
    \caption{The distribution of T24-Deep Ly$\alpha$ emitting galaxies in $\Delta v$ and ${\rm M}_{\rm UV}$ (black), with the size of the outlier MUSE-1670 exaggerated for emphasis. The mean value of $\Delta v({\rm M}_{\rm UV})$ predicted by our model, with $1\sigma$ bounds, is overlaid in red. A naive relation obtained by linear regression is also overlaid in blue. The right panel shows the probability density functions of our model and the naive model at ${\rm M}_{\rm UV}=-21.3$, the ${\rm M}_{\rm UV}$ of MUSE-1670. Our model clearly indicates that this object is an outlier, while the naive model does not.}
    \label{fig:outlier}
\end{figure}

\begin{figure*}
    \resizebox{\hsize}{!}
    {\includegraphics[width=\textwidth]{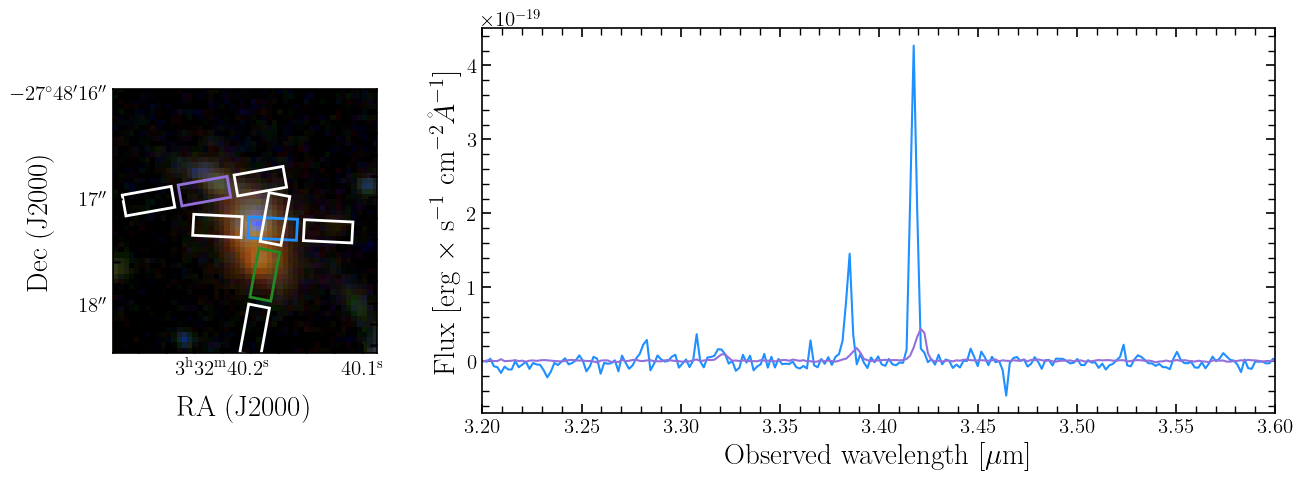}}
    \caption{The MUSE-1670 object with \textit{JWST} pointings overlaid (left panel) and blue and purple spectra (right panel), taken from the JADES DR3 data release \citep{deugenio25}. The green spectrum is not included as it is not yet publicly available, but will be made so in the next JADES data release. The blue, purple, and green spectra all include H$\alpha$ at systemic redshifts of $5.8282$, $5.8312$, and $5.8193$.}
    \label{fig:muse1670}
\end{figure*}

While most galaxies in the T24 sample lie well within the $95\%$ probability density region predicted by our model (see Figure \ref{fig:hist}), we note one significant outlier: MUSE-1670 (located in the MUSE-Deep field, the galaxy with the highest $\Delta v$ in the top middle panel of Figure \ref{fig:hist}).  

The identification of outliers in a dataset depends on the assumed distribution of the data, which can vary greatly depending on how observational effects are accounted for. 
We illustrate this using a scatterplot of $\Delta v$ and ${\rm M}_{\rm UV}$ values of galaxies in the T24-Deep dataset in Figure \ref{fig:outlier}. In red we overlay the $\Delta v-{\rm M}_{\rm UV}$ found by our model with $1\sigma$ scatter (c.f. Eq. \ref{eq:dv}) and in blue we overlay a linear regression to the data whose scatter is equal to the root mean squared difference between the data and the regression. The right panel shows the conditional probability distributions $P(\Delta v|{\rm M}_{\rm UV}=-21.3)$ for each model with the MUSE-1670 value indicated with the dashed line. Our model assigns a much lower probability density to the value of $\Delta v$ measured for MUSE-1670, 
  than the linear regression model. As such, our model allows us to accurately flag MUSE-1670 as an outlier, which would have otherwise been missed using the linear regression model.

In investigating this anomalous object, we noted a cursory similarity between its Ly$\alpha$ emission properties and those of the LAE GN-z11 \citep{bunker23}. These properties, as well as the conditional probability distributions for $P(W_{\rm emerg}^{\rm Ly\alpha},\Delta v,f_{\rm esc}^{\rm Ly\alpha}|{\rm M}_{\rm UV}=-21.5)$ predicted by our model, are listed in Table \ref{tab:analog}. The escape fraction listed for our model assumes case A recombination: the value assuming case B recombination is $3\%$ lower.

While both MUSE-1670 and GN-z11 have $f_{\rm esc}^{\rm Ly\alpha}$ and $W_{\rm emerg}^{\rm Ly\alpha}$ typical of ${\rm M}_{\rm UV}=-21.5$ galaxies, their $\Delta v$ lie $2.12\sigma$ and $3.01\sigma$ above the modal value of our model, respectively. 
MUSE-1670 does not appear in any radio galaxy or active galactic nuclei (AGN) catalogs, nor does it feature broad line emission, meaning it is unlikely to be an anomalous AGN which survived the AGN exclusion selection of MUSE-Deep \citep{bacon17}. The JADES survey in the GOODS-S field targeted this galaxy with three separate pointings over the surface area of MUSE-1670\footnote{MUSE-1670 may be viewed at this URL: \url{https://s3.amazonaws.com/grizli-v2/ClusterTiles/Map/gds/jwst.html?coord=53.1666918,-27.8042720&zoom=8}}, which is extended over $\sim5.9$ proper kpc. Figure \ref{fig:muse1670} shows MUSE-1670, along with the overlapping \textit{JWST} pointings and measured spectra. The spectra yield slightly different systemic redshifts: $5.8282$, $5.8312$, and $5.8193$ for the pointings colored purple, blue, and green in Figure \ref{fig:muse1670}, respectively. These correspond to velocity offsets of $2668$ and $3573$ km s$^{-1}$ for the purple and blue spectra with respect to the rest frame of the green spectrum. The large spatial extent of MUSE-1670, different colors of its spatial components, and the vast range of systemic velocities support the interpretation that MUSE-1670 corresponds to a merging system. Interestingly, GN-z11 is similarly compact, with a proper size of $1.2$ kpc \citep{bunker23}. If the Ly$\alpha$ properties of MUSE-1670 are typical of a merger-induced starburst, then we may posit that GN-z11 is also undergoing a merger.

\subsection{Redshift evolution}

Our model assumes that the probability distribution of emergent Ly$\alpha$ properties is conditional {\it only} on ${\rm M}_{\rm UV}$, and therefore its redshift evolution $dp(x_\alpha)/dz$ is simply $dp(x_\alpha)/d{\rm M}_{\rm UV} \times d{\rm M}_{\rm UV}/dz$. We expect this to be a reasonable approximation for $L_{\rm Ly\alpha}$ and $L_{\rm H\alpha}$, which both scale with the SFR that drives ${\rm M}_{\rm UV}$ (e.g., \citealt{kennicut98,sobral19}), albeit with with the caveat that $L_{\rm Ly\alpha}$ and ${\rm M}_{\rm UV}$ trace the SFR on $10$ Myr and $100$ Myr timescales, respectively, so any redshift evolution in the relationship between SFRs on these timescales (e.g., higher burstiness at earlier redshifts, c.f. \citealt{sun25b}) could introduce shifts in our inferred relations at $z>5-6$. Our approximation is unlikely to hold as well for $\Delta v$, which is thought to more directly trace the gas mass \citep{stark17,hayes23}. Including explicit redshift dependence in $P(\Delta v)$ may increase the likelihood of observing objects like GS-z13 and GN-z11. In future work we will explore the physical causes of redshift evolution in Ly$\alpha$ observables and their impact on inferences of reionization.

\section{Conclusion}
\label{sec:concl}


We present a new empirical model of emergent Ly$\alpha$ emission intended for use in EoR studies. Our model is calibrated to the set of MUSE LAEs presented in \cite{tang24} which also have detected H$\alpha$ from \textit{JWST}. In our calibration, we forward-model the selection effects of both the MUSE and \textit{JWST} surveys; this is crucial in avoiding spurious correlations induced by the selection. The resulting model is a conditional multivariate Gaussian described in Section \ref{sec:quick}. We list the defining characteristics of our model below.

\begin{itemize}
    \item The distribution of $\log_{10}L_{\rm Ly\alpha}$, $\Delta v$, and $\log_{10}L_{\rm H\alpha}$ is a multivariate Gaussian whose mean varies with ${\rm M}_{\rm UV}$ and whose covariance matrix is constant across ${\rm M}_{\rm UV}$.
    \item The inclusion of correlations between Ly$\alpha$ properties allows us to estimate conditional probabilities, such as $P(W_{\rm emerg}^{\rm Ly\alpha}|\Delta v)$, which is useful in cases where one property is measured while only a limit is available for another.
    \item The model calibration accounts for the expected number of LAEs in the survey area as a function of ${\rm M}_{\rm UV}$, enabling the model to reproduce the Ly$\alpha$ luminosity function without calibrating to additional data.
    \item Forward modeling of selection effects during calibration ensure that the covariances present in our model reflect the true distribution of the data.
\end{itemize}

Our model accurately reproduces the equivalent width probability density function and Ly$\alpha$ luminosity function of \cite{umeda25}, even without additional calibration, as seen in Figures \ref{fig:ewpdf} and \ref{fig:laelf}. It also predicts several scalings for Ly$\alpha$ parameters (Eq.s \ref{eq:lya}-\ref{eq:fesc}), reproducing the findings of \cite{mason2018} and \cite{prietolyon25} that low-UV luminosity galaxies tend to have higher $W_{\rm emerg}^{\rm Ly\alpha}$ and lower $\Delta v$. The principal physical implications of our model are listed below.

\begin{itemize}
    \item Ly$\alpha$ equivalent width and escape fraction tend to increase with decreasing UV luminosity.
    \item The Ly$\alpha$ red wing velocity offset tends to increase with increasing UV luminosity.
    \item The only significant outlier in our calibration data, MUSE-1670, is likely undergoing a merger. This outlier incidentally has Ly$\alpha$ and galaxy characteristics similar to those of GN-z11, lending credence to the idea that GN-z11 may be a merger.
    \item Compared to some previous works, our model predicts a more extended distribution of equivalent widths and Ly$\alpha$ velocity offsets, which could facilitate the detection of Lyman alpha in the early EoR stages.
    \item We provide fitting functions in Eq.s \ref{eq:lya}-\ref{eq:ew_fit} and make our code publicly available\footnote{\url{https://github.com/samgagnon/empirical_lya}}.
\end{itemize}

Since our model has a clear separation of galaxy properties and survey parameters, one can include additional datasets with corresponding selection functions, thereby improving the inferred constraints on galaxy properties.  With this new characterization of emergent Ly$\alpha$ from high-$z$ galaxies, we will next try to robustly isolate EoR damping wing imprints from {\it JWST} and Subaru observations.

\begin{acknowledgements}
    We thank M. Wyatt, C. Mason, D. Stark for insightful comments on a draft version of this work.  We extend further gratitude to B. Trefoloni, S. Shanbhog, and S. Carniani for useful discussion on observational effects.
    We gratefully acknowledge computational resources of the HPC center at SNS.
    AM acknowledges support from the Italian Ministry of Universities and Research (MUR) 
    through the PRIN project ``Optimal inference from radio images of the epoch of 
    reionization'', and the PNRR project ``Centro Nazionale di Ricerca in High Performance 
    Computing, Big Data e Quantum Computing''.
\end{acknowledgements}

\bibliography{bibliography}{}
\bibliographystyle{aa}

\begin{onecolumn}

\begin{appendix}

\section{SO(3) Lie group}
\label{apx:matrices}

The basis space of the change of basis matrix $A$ spans the space of all 3D rotations plus the space of all scalar multiples via the identity matrix \citep{fletcher03}. While the full set of 3D basis transformations also includes stretching and shearing, the matrix $A$ operates on a quantity which is already normalized to have unit variance, and should therefore not be related to its orthonormal basis by a stretch transformation. Furthermore, our basis can still produce the subset of shear transformations which lie in the convex cone generated by $\mathcal{I}$ and SO(3) \citep{Chirikjian2011}. While this does not cover all possible shear transformations, we assume that it contains at a minimum one of the shear transformations required to orthonormalize the T24 data. Since our MLE converged on a convincing solution, we deem these assumptions reasonable.

The transformation matrices of the SO(3) Lie group are 

\begin{equation}
    S_1=
    \begin{bmatrix}
        0 & 0 & 0 \\
        0 & 0 & -1 \\
        0 & 1 & 0 \\
    \end{bmatrix}, \;
    S_2=
    \begin{bmatrix}
        0 & 0 & 1 \\
        0 & 0 & 0 \\
        -1 & 0 & 0 \\
    \end{bmatrix}, \; {\rm and} \;
    S_3=
    \begin{bmatrix}
        0 & -1 & 0 \\
        1 & 0 & 0 \\
        0 & 0 & 0 \\
    \end{bmatrix}.
\end{equation}

\noindent Linear combinations of SO(3) rotation matrices are themselves also rotation matrices, therefore preserving the origin about which vectors are rotated and the magnitude of the vector. By defining our change of basis matrix $A$ as a linear combination of SO(3) rotation group matrices and the identity matrix, we impose that the data vector $x_{\alpha}$ is a linear combination of $u$ with a 3D rotation of $u$. 

\section{Gaussian likelihood function}
\label{appdx:like2}

In the second step of our model calibration procedure, we minimize the objective function

\begin{equation}
    {\rm obj}=\Sigma^i \Sigma_{j}\left(\frac{\langle y\rangle_j^i-\mu_j^i}{\sigma_j^i}\right)^2,
\end{equation}

\noindent where $i$ represents the index of each dataset, $j$ represents the index of each quantity computed from our model and the data, $\langle y\rangle_j^i$ is the expectation value of quantity $j$ assuming the selection function of $i$ computed from our model, $\mu_j^i$ is the same quantity computed from the data, and $\sigma_j^i$ is the standard deviation of that quantity in the data. We use

\begin{equation}
    i\in\{\rm T24-Wide,T24-Deep\}, \;
    j\in\{f_{\rm obs}({\rm M}_{\rm UV}),W_{\rm emerg}^{\rm Ly\alpha}({\rm M}_{\rm UV}),\Delta v({\rm M}_{\rm UV}),f_{\rm esc}^{\rm Ly\alpha}({\rm M}_{\rm UV})\},
\end{equation}

\noindent where each quantity in $j$ is computed for $20$ equal bins in ${\rm M}_{\rm UV}$ ranging from $-20$ to $-17$.

\section{Model parameters}
\label{apdx:params}

\begin{table*}
    \centering
    \caption{All model parameters, including the selection criteria with prior ranges.}
    \begin{tabular}{c|c|c|c}
        \hline
         Parameter & Description & Prior range & Optimal value \\
         \hline
         $c_1$ & \tiny transformation coefficient 1 & $[-1, 1]$ & $1.0$ \\
         $c_2$ & \tiny transformation coefficient 2 & $[-1, 1]$ & $1.0$ \\
         $c_3$ & \tiny transformation coefficient 3 & $[-1, 1]$ & $0.34$ \\
         $c_4$ & \tiny transformation coefficient 4 & $[-1, 1]$ & $-1.0$ \\
         $m_1$ & \tiny slope of mean of eigenvector 1  & $[-1, 1]$ & $8.7\cdot10^{-2}$ \\
         $m_2$ & \tiny slope of mean of eigenvector 2  & $[-1, 1]$ & $-0.57$ \\
         $m_3$ & \tiny slope of mean of eigenvector 3  & $[-1, 1]$ & $-0.38$ \\
         $b_1$ & \tiny intercept of mean of eigenvector 1  & $[-3, 3]$ & $-0.51$ \\
         $b_2$ & \tiny intercept of mean of eigenvector 2  & $[-3, 3]$ & $-0.85$ \\
         $b_3$ & \tiny intercept of mean of eigenvector 3  & $[-3, 3]$ & $-0.31$ \\
         $\sigma_1$ & \tiny standard deviation of eigenvector 1  & $[0.01, 1.0]$ & $0.70$ \\
         $\sigma_2$ & \tiny standard deviation of eigenvector 2  & $[0.01, 1.0]$ & $0.49$ \\
         $\sigma_3$ & \tiny standard deviation of eigenvector 3  & $[0.01, 1.0]$ & $0.26$ \\
         $W_{\rm Wide}$ & \tiny{MUSE-Wide Ly$\alpha$ EW limit} & $[40,120]$ & $116$\\
         $W_{\rm Deep}$ & \tiny{MUSE-Deep Ly$\alpha$ EW limit} & $[12.5,50]$ & $24$\\
         $f_{\rm Wide}$ & \tiny{MUSE-Wide Ly$\alpha$ flux limit} & $[10^{-17},3\cdot10^{-17}]$ & $1.8\cdot10^{-17}$\\
         $f_{\rm Deep}$ & \tiny{MUSE-Deep Ly$\alpha$ flux limit} & $[10^{-18},3\cdot10^{-18}]$ & $2.7\cdot10^{-18}$\\
         $f_{\rm \it JWST}$ & \tiny{{\it JWST} H$\alpha$ flux limit} & $[10^{-18},3\cdot10^{-18}]$ & $1.2\cdot10^{-18}$\\
         $\mu_{\rm Ly\alpha}$ & \tiny mean of data in MUSE-Deep field & - & $\{42.47,\;200.18,\;42.03\}$ \\
         $\sigma_{\rm Ly\alpha}$ & \tiny standard deviation of data in MUSE-Deep field & - & $\{0.42,\;99.7,\;0.39\}$ \\
         \hline
    \end{tabular}
    \label{tab:params}
\end{table*}

Table \ref{tab:params} lists all parameters of our model.

\section{Multivariate normal parameterization of model}
\label{apdx:mult_norm}

The model of $P(\log_{10}L_{\rm Ly\alpha},\Delta v,\log_{10}L_{\rm H\alpha}|{\rm M}_{\rm UV})$ presented in this work may be written in closed form as a multivariate normal distribution

\begin{equation}
    \begin{bmatrix}
        \log_{10}L_{\rm Ly\alpha}\\ \Delta v \\ \log_{10}L_{\rm H\alpha}
    \end{bmatrix}=\mathcal{N}\left(
    \begin{bmatrix}
        -0.28({\rm M}_{\rm UV}+18.5)+41.86{\;\rm erg\;s^{-1}\;cm^{-2}} \\ 
        -27.92({\rm M}_{\rm UV}+18.5)+197.19{\;\rm km\;s^{-1}}\\
        -0.38({\rm M}_{\rm UV}+18.5)+41.64{\;\rm erg\;s^{-1}\;cm^{-2}}\\
    \end{bmatrix},
    \begin{bmatrix}
        0.13 & -0.11 & 0.02 \\
        -0.11 & 7998.70 & 13.26 \\
        0.02 & 13.26 & 0.06 \\
    \end{bmatrix}
    \right),
\end{equation}

\noindent where the entries of $\mathcal{N}$ are its mean, $\mu$, and covariance, $\Sigma$.

\section{Equivalent width PDF}
\label{apdx:ewpdf}

\begin{figure*}
    \resizebox{\hsize}{!}
    {\includegraphics[width=\textwidth]{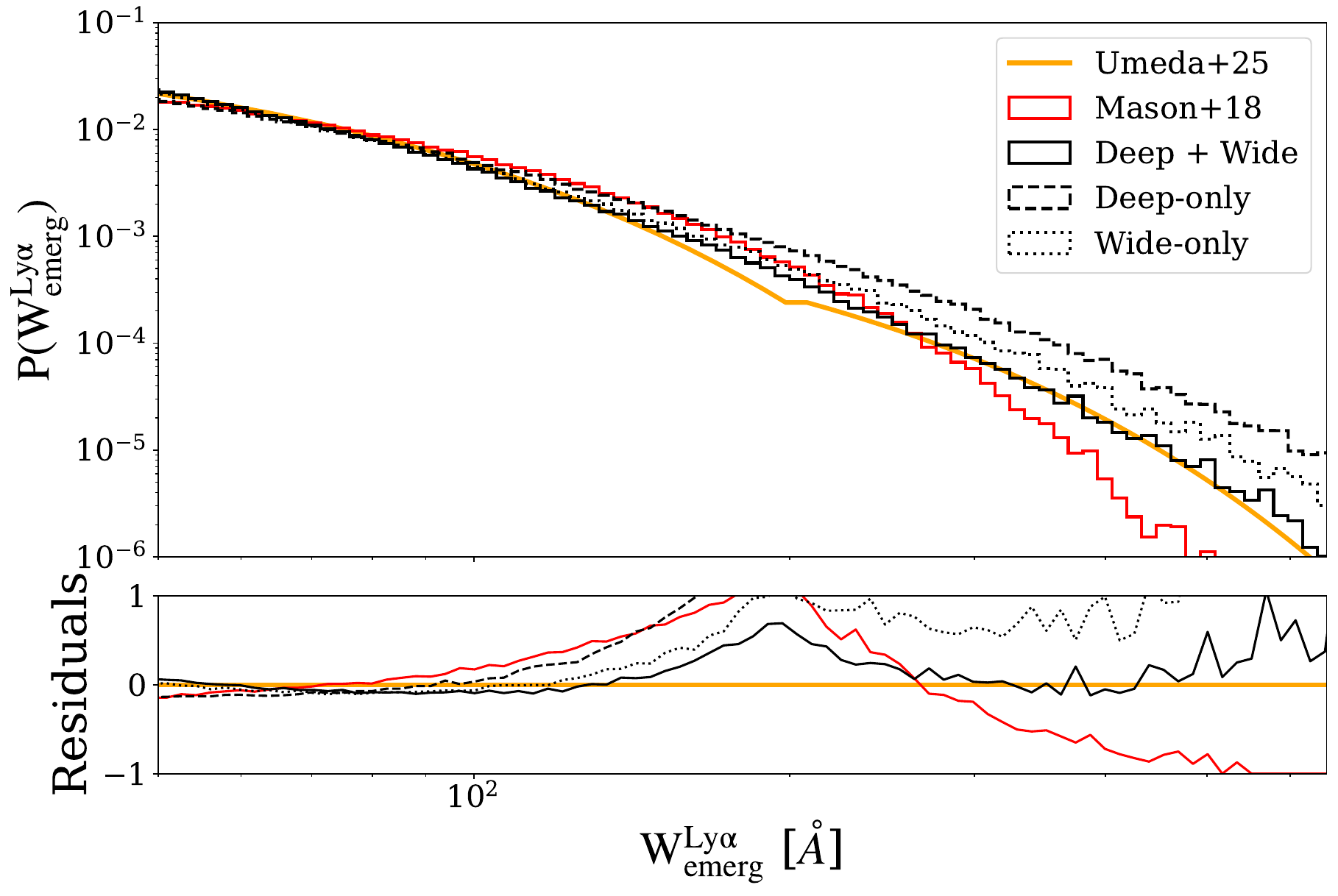}}
    \caption{The distribution of equivalent widths for $z\sim5$ galaxies with ${\rm M}_{\rm UV}\leq-18$ measured by \cite{umeda25} (orange) and predicted by the models of \cite{mason2018} (red) and this work (black). The dashed and dotted curves show the fits which would be obtained if we used only the T24-Deep and T24-Wide datasets (see Table \ref{tab:datasets}), respectively, in our model calibration. The bottom panel shows residuals against the measured relation.}
    \label{fig:ew_res}
\end{figure*}

Figure \ref{fig:ew_res} shows the distribution of Ly$\alpha$ equivalent widths (EWPDF) for galaxies at $z\sim5$ with ${\rm M}_{\rm UV}\leq-18$ measured by \cite{umeda25} and predicted by the models of \cite{mason2018} (M18) and this work. We posit that our model can better reproduce the measured EWPDF because it is calibrated on both a deep (long exposure, small angular area) and a wide (short exposure, large angular area) dataset. The former should be dominated by dimmer, more numerous objects, while the latter should be brighter, rarer objects. These correspond to the peak and faint tail of the EWPDF, respectively.

To investigate this hypothesis, we recalibrated our model using only the T24-Deep dataset, and then again using only the T24-Wide dataset (see Table \ref{tab:datasets}). The EWPDFs arising from these models are shown in dashed and dotted curves in Figure \ref{fig:ew_res}. The bottom panel includes the residuals of each model with respect to the measured EWPDF of \cite{umeda25}. Note that while the Deep-only model diverges significantly from all other models at high equivalent widths, it converges to the M18 model at low equivalent widths. This makes sense since the calibration sample of the M18 model were rare faint objects similar to those in the T24-Deep sample. Meanwhile, the Wide-only model more accurately fits the measured relation up to an equivalent width of $\sim200$, but then diverges. The measured relation is only convincingly recovered when both datasets are included.

It is possible that the variance in the ${\rm M}_{\rm UV}-{\rm W}_{\rm emerg}^{\rm Ly\alpha}$ relation dominates the probability density at the faint end of the EWPDF. In this case, the model calibrated to both datasets is better able to recover that faint end thanks to its accurate determination of the variance in the ${\rm M}_{\rm UV}-{\rm W}_{\rm emerg}^{\rm Ly\alpha}$ relation, constrained by fitting the observed fractions shown in Figure \ref{fig:fobs}.

\section{M18 recalibration}
\label{apdx:m18}

The recalibrated M18-like relations for the probability of emitting Ly$\alpha$ with $f_{\rm Ly\alpha}\geq2\cdot 10^{-18}$ erg s$^{-1}$ is $A({\rm M}_{\rm UV})$ and the mean equivalent width is $W_c({\rm M}_{\rm UV})$ are:

\begin{equation}
    A({\rm M}_{\rm UV})=0.50-0.50\tanh\left[0.67({\rm M}_{\rm UV}+18.0)\right],
    \label{eq:a_m18}
\end{equation}
\noindent and 
\begin{equation}
    W_c({\rm M}_{\rm UV})=13.95+169.09\exp\left[0.64({\rm M}_{\rm UV}+16.5)\right]\;\AA.
    \label{eq:wc_m18}
\end{equation}

Our model assumes that all galaxies emit in Ly$\alpha$. For the purposes of recalibrating the M18 model, we compute $A({\rm M}_{\rm UV})$ as the fraction of emitters with $f_{\rm Ly\alpha}\geq2\cdot10^{-18}$ erg s$^{-1}$, as this was the $5\sigma$ detection threshold of the VANDELS survey \citep{pentericci18}. The $\tanh$ function reported for $A({\rm M}_{\rm UV})$ in Eq. \ref{eq:a_m18} fits our model extremely well. This makes sense since we model $P(\log_{10}L_{\rm Ly\alpha}|{\rm M}_{\rm UV})$ as a Gaussian distribution, whose cumulative distribution function, the error function, is very similar in form to the $\tanh$ function. We fit $W_c({\rm M}_{\rm UV})$ to the mean equivalent width assuming $f_{\rm Ly\alpha}\geq2\cdot10^{-18}$, i.e. $\mathbb{E}\{P(W_{\rm Ly\alpha}|{\rm M}_{\rm UV},f_{\rm Ly\alpha}\geq2\cdot10^{-18})\}$. Here the $\tanh$ function is less appropriate. While the best-fit $\tanh$ holds well for ${\rm M}_{\rm UV}\approx18$ and fainter, its error exceeds $5\%$ at ${\rm M}_{\rm UV}=-21$ and reaches $30\%$ at ${\rm M}_{\rm UV}=-22$. Switching to an exponential distribution reduces all errors in the range ${\rm M}_{\rm UV}\in[-22,\;-16]$ to $<5\%$.

\section{$M_h-{\rm M}_{\rm UV}$ relation}
\label{appdx:mh_muv}

In our investigation of the $\Delta v-{\rm M}_{\rm UV}$ relation predicted by our model, we employed the ${\rm M}_{\rm UV}-M_h$ relation of \cite{davies25}. In that work, ${\rm M}_{\rm UV}$ is treated as a function of the star formation rate, following the implementation of \cite{park2019}:

\begin{equation}
    {\rm M}_{\rm UV}=51.6-2.5\log_{10}\left(\frac{\rm SFR}{\kappa_{\rm UV}}\right),
\end{equation}

\noindent where $\kappa_{\rm UV}=1.15\cdot10^{-28}$ M$_\odot$ yr$^{-1}$ erg$^{-1}$ s Hz, following \cite{sun16}. \cite{davies25} samples SFR from the log-normal distribution:

\begin{equation}
    P(\ln({\rm SFR})|M_*)=\mathcal{N}(\mu_{\rm SFR}(M_*),\sigma_{\rm SFR}(M_*)),
\end{equation}

\noindent where $M_*$ is the stellar mass, and $\mu_{\rm SFR}$ and $\sigma_{\rm SFR}$ are functions of $M_*$. The mean follows the relation

\begin{equation}
    \mu_*(M_*)=\frac{M_*}{t_*H(z)}
\end{equation}

\noindent from \cite{park2019}, where $t_*=0.5$ and $H(z)$ is the Hubble constant at redshift $z$. The standard deviation increases with increasing stellar mass according to a double power-law:

\begin{equation}
    \sigma_{\rm SFR}={\rm max}\left(\sigma_{\rm SFR,lim},\sigma_{\rm SFR,idx}\log_{10}\left(\frac{M_*}{10^{10}\;{\rm M}_{\odot}}\right)+\sigma_{\rm SFR,lim}\right),
\end{equation}

\noindent where $\sigma_{\rm SFR,lim}=0.19$ and $\sigma_{\rm SFR,idx}=-0.12$. Similarly, \cite{davies25} sample $M_*$ from the log-normal distribution

\begin{equation}
    P(\ln M_*|M_h)=\mathcal{N}(\mu_*(M_h),\sigma_*),
\end{equation}

\noindent where $\sigma_*=0.5$ and $\mu_*(M_h)$ is the scaling relation of \cite{mirocha2017}:

\begin{equation}
    \mu_*(M_h)=f_{*,10}\left(\frac{(M_{\rm pivot}/10^{10}\;{\rm M}_\odot)^{\alpha_*}+(M_{\rm pivot}/10^{10}\;{\rm M}_\odot)^{\alpha_{*2}}}{(M_h/10^{10}\;{\rm M}_\odot)^{-\alpha_*}+(M_h/10^{10}\;{\rm M}_\odot)^{-\alpha_{*2}}}\right)M_h\exp\left(-\frac{M_{\rm turn}}{M_h}\right)\frac{\Omega_b}{\Omega_m},
\end{equation}

\noindent where $M_{\rm pivot}=2.8\cdot10^{11}$ M$_\odot$, $\alpha_*=0.5$, $\alpha_{*2}=-0.61$, and $M_{\rm turn}=5\cdot10^8$ M$_\odot$.

Using the equations above, we computed $P({\rm M}_{\rm UV}|M_h)$ for $M_h\in[10^8\;{\rm M}_\odot,\;10^{12}\;{\rm M}_\odot]$ and inverted this distribution numerically to produce $P(M_h|{\rm M}_{\rm UV})$.

\section{Conditional probabilities of the multivariate normal distribution}
\label{appdx:conditionals}

Let $x$ follow a multivariate normal distribution $x\sim\mathcal{N}(\mu,\Sigma)$, characterized by its mean, $\mu$, and covariance matrix $\Sigma$, with components

\begin{equation}
    \mu=
    \begin{bmatrix}
        \mu_1 \\ \mu_2 \\ \mu_3
    \end{bmatrix},
\end{equation}

\noindent and

\begin{equation}
    \Sigma=
    \begin{bmatrix}
        \Sigma_{11} & \Sigma_{12} \\
        \Sigma_{21} & \Sigma_{22} \\
    \end{bmatrix}.
\end{equation}

\noindent The probability distribution of $x_1$ conditioned on a value of $x_2$ drawn from the same distribution is

\begin{equation}
    P(x_1|x_2)=\mathcal{N}(\mu_{1|2},\Sigma_{1|2}),
\end{equation}

\noindent where

\begin{equation}
    \mu_{1|2}=\mu_1+\Sigma_{12}\Sigma_{22}^{-1}(x_2-\mu_2)
\end{equation}

\noindent and

\begin{equation}
    \Sigma_{1|2}=\Sigma_1-\Sigma_{12}\Sigma_{22}^{-1}\Sigma_{21}.
\end{equation}

Using these expressions, we may obtain a closed form also for $P(x_1|x_2>X_2)$, where $X_2$ is some fixed minimum value of $x_2$. This is

\begin{equation}
    P(x_1|x_2>X_2)\propto P(x_1)\int_{X_2}^{\infty}P(x_2|x_1)dx_2,
\end{equation}

\noindent which, using the cumulative density function of the normal distribution, $\Phi(X;\mu,\Sigma)$, is

\begin{equation}
    P(x_1|x_2>X_2)\propto P(x_1)\left[1-\Phi(X_2;\mu_{2|1},\Sigma_{2|1})\right].
\end{equation}

\noindent Similarly, the probability density function of $x_1$ given a maximum value of $x_2$ is

\begin{equation}
    P(x_1|x_2<X_2)\propto P(x_1)\Phi(X_2;\mu_{2|1},\Sigma_{2|1}).
\end{equation}

\end{appendix}

\end{onecolumn}

\end{document}